# Ambipolar metal–insulator transition in bilayer $MoSe_2$

Ha-Leem Kim[1,2,3,*], Hyungbin Lim[1,2], Terry Wu[1], Amogh Y. Waghmare[1], Garima Gupta[4], Takashi Taniguchi[5], Kenji Watanabe[6], Archana Raja[4], Eric Y. Ma[1,2,7], Feng Wang[1,2,3,*]

[1]Department of Physics, University of California at Berkeley, Berkeley, CA, USA.
[2]Materials Sciences Division, Lawrence Berkeley National Laboratory, Berkeley, CA, USA.
[3]Kavli Energy NanoScience Institute, University of California at Berkeley and Lawrence Berkeley National Laboratory, Berkeley, CA, USA.
[4]Molecular Foundry, Lawrence Berkeley National Laboratory, Berkeley, CA, USA.
[5]Research Center for Materials Nanoarchitectonics, National Institute for Materials Science, Tsukuba, Japan.
[6]Research Center for Electronic and Optical Materials, National Institute for Materials Science, Tsukuba, Japan.
[7]Department of Electrical Engineering and Computer Sciences, University of California at Berkeley, Berkeley, CA, USA.

*Corresponding author. Email: fengwang76@berkeley.edu (F.W.); haleem4644@berkeley.edu (H.-L.K.)

## Abstract

When Coulomb interactions dominate kinetic energy, dilute two-dimensional carriers form a Wigner solid. Both its melting and disorder-driven localization can drive density-tuned metal–insulator transitions, but their relative roles and interplay have been debated for decades. Atomically thin semiconductors offer a new test because transport and scanning tunneling microscopy can probe the same material, yet poor contacts impede transport where correlations are strongest. Here we introduce contact-free radio-frequency reflectometry that senses sheet resistance through capacitive gate coupling. It resolves the electron and previously inaccessible hole transitions in the same bilayer $MoSe_2$ device, whose carriers share disorder but not mass. Their critical densities differ fourfold yet correspond to comparable Coulomb-to-kinetic-energy ratios, where tunneling microscopy finds solid–liquid coexistence. Magnetoresistance peaks sharply at each transition, reflecting a Pomeranchuk-like, field-driven expansion of the Wigner solid against the correlated liquid. Our results demonstrate interaction-driven transitions in high-quality two-dimensional semiconductors, opening quantitative transport in contact-limited systems.

The metal–insulator transition (MIT) of interacting electrons in two dimensions is a long-standing problem in condensed-matter physics(*1-3*). When Coulomb interactions dominate kinetic energy, electrons can crystallize into a Wigner crystal(*4, 5*). The balance between these energies is characterized by the Wigner–Seitz radius $r_s = a_0/a_B^* = m^* e^2/(4\pi\epsilon\hbar^2\sqrt{(\pi n)})$, where $a_0$ is the average interparticle distance, $a_B^*$ is the effective Bohr radius, $n$ is the carrier density, $\epsilon$ is the effective dielectric constant, and $m^*$ is the effective mass. This parameter increases as the carrier density is reduced, with clean-limit calculations placing the crystallization threshold at $r_s \gtrsim 30$(*4, 6-8*). Near a putative first-order transition between the correlated liquid and the Wigner crystal, long-range Coulomb interactions frustrate macroscopic phase separation and can favor mesoscopic mixtures of the two phases(*9, 10*). How the Wigner solid melts in the presence of disorder and whether this melting underlies the observed two-dimensional MIT remain debated.

Transport studies in two-dimensional semiconductor systems, most extensively silicon MOSFETs and GaAs heterostructures, have revealed a rich set of phenomena near the MIT and motivated both interaction- and disorder-based descriptions(*11-20*). A central experimental challenge is to determine what sets the critical density and how the interplay between interactions and disorder controls transport near the transition. Atomically thin semiconductors have emerged as a powerful platform for addressing these questions(*19, 21, 22*). Different transition-metal dichalcogenide (TMD) layers can have different electron and hole masses, placing the strongly correlated regime at various densities and temperatures(*23-27*). In addition, the exposed TMD layers allow correlated electronic states to be imaged directly by scanning tunneling microscopy (STM)(*28-32*). Transport can determine critical densities and systematically map the temperature- and magnetic-field-dependent conduction for both electrons and holes, whereas STM directly resolves the local electronic phases and disorder landscape. Applying both approaches to the same material can therefore connect the transport signatures of the MIT to the microscopic states producing them.

STM has imaged quantum melting behavior of Wigner solids in several 2D semiconductors, including both electrons and holes in bilayer MoSe2(*29, 31*). Corresponding ambipolar transport measurements could shed new light on the MIT mechanism: electrons and holes in the same sample share the underlying disorder landscape but can have markedly different masses and band degeneracies, so the ordering of their critical densities $n_c$ can discriminate between interaction- and disorder-dominating criteria. The limitation posed by the electrical contacts, however, hampers electrical transport measurements in many 2D semiconductors, particularly for low-density carriers with a large effective mass(*25, 33, 34*).

Here we realize ambipolar transport measurements of MIT transitions in bilayer $MoSe_2$ using a contact-free radio-frequency (RF) reflectometry readout of the in-plane channel resistance. The RF voltage drives the carrier transport through capacitive coupling, and its response is read out from the gate admittance. Bilayer MoSe2 provides an attractive model system. In addition to being able to compare with the recent STM imaging data, bilayer $MoSe_2$ has distinctly different electron and hole bands — the valence-band maximum at Γ has an effective mass $m_h^* \approx 1.26\ m_e$ and a twofold spin degeneracy, whereas the conduction-band minima at the Q valleys have $m_e^* \approx 0.54\ m_e$ and a total degeneracy of twelve(*29, 35*). We find that the electron and hole MIT transitions occur at comparable Wigner–Seitz radii, $r_s \approx 18.0$ for electrons and $r_s \approx 21.7$ for holes, despite the hole mass being more than two times larger. In a similar $r_s$ range, STM imaging observed mixed Wigner solid and liquid phases in bilayer $MoSe_2$. A common Ioffe–Regel–Mott (IRM) crossover criterion, evaluated using the measured mobilities and band degeneracies, instead predicts a lower

MIT crossover density for holes than for electrons. This contrast supports interaction strength as the key factor anchoring the MIT transition in high-quality 2D semiconductors. Magnetotransport further reveals a pronounced magnetoresistance (MR) peak sharply localized at each carrier's critical density. We interpret this as arising from a magnetic-field-induced change in the balance between Wigner-solid (Curie-like spin) and correlated-liquid (Pauli-like spin) regions, with the resulting resistance change amplified as the percolating liquid network approaches its connectivity threshold near the transition.

**Contact-free radio-frequency readout of the channel resistance**

The RF device consists of a 2L-$MoSe_2$ channel encapsulated in hBN, contacted at one end by a graphite electrode (Fig. 1, A and B). A metallic top-gate covers roughly half of the channel, dividing it into two regions in series: Region 1, gated by the back gate alone and adjoining the graphite contact, and Region 2, gated by both the top and the back-gate. A separate contact gate heavily dopes the channel near the graphite contact to maintain rapid carrier equilibration during gate sweeps. The top-gate is connected to a 50-Ω line through a bias tee and an impedance-matching network [(*36*), section 2]. The matching inductance and the total shunt capacitance at the device node together set the resonance frequency, which lies in the 180–200 MHz range.

At the operating frequency, the impedance of the graphite–$MoSe_2$ contact is large, so negligible AC current flows through it. Instead, the RF excitation drives carriers back and forth between Region 1 and Region 2 across their boundary (red arrow in Fig. 1A): the AC current enters the channel capacitively through the top hBN, flows laterally along the channel, and returns capacitively through the bottom hBN to the AC-grounded back-gate. The resonator thus senses the in-plane motion of carriers directly. The RF power applied to the 50 Ω input port feeding the matching-network/sample assembly was calibrated to $P_{in}$ = −92 dBm; power-dependence tests and distributed-RC simulations indicate that the RF readout is non-perturbative under these conditions [(*36*), sections 1 and 3].

Figure 1, C and D, shows the reflected amplitude $|S_{11}|$ (C) and phase (D) spectra as a function of back-gate voltage $V_{BG}$ at $T = 2$ K, measured separately for electron and hole doping. The top-gate DC bias $V_{TG}$ is set so that both regions reach charge neutrality at the same $V_{BG}$, keeping the two regions density matched throughout the sweep [(*36*), section 4]. We quantify this gate-dependent evolution with a global line-shape fit that shares a single set of background parameters across the gate sweep [(*36*), section 5]; the extracted $f_0$ and $\min|S_{11}|$ are shown in Fig. 1E, with the conduction onset marked by dashed lines. At charge neutrality, the channel is insulating and the resonance sits at its off-state value $f_{off}$; as $V_{BG}$ dopes the channel, $f_0$ decreases monotonically and $\min|S_{11}|$ tracks how closely the device impedance matches the 50-Ω line.

This behavior follows from the resistance dependence of RF charge-penetration length. At low sheet resistance $R$, the RF penetration length exceeds the length of the back-gate-only region, so both Region 1 and Region 2 participate in the gate admittance. At high $R$, the RF-driven charge redistribution penetrates only a short distance into Region 1 from the Region 1–Region 2 boundary, while Region 2 remains capacitively coupled to the top-gate along its full length. The effective device capacitance therefore decreases as the Region 1 channel becomes resistively decoupled, shifting the resonance from the conducting-limit value $f_{on}$ toward the insulating-limit value $f_{off}$. More quantitatively, for a uniform channel section, the penetration length is $\delta = \sqrt{(2/(\omega R C_A))}$, where $C_A$ is the relevant gate capacitance per unit area. The frequency response is steepest when $\delta$ is comparable to the corresponding region length $L$, giving the approximate resistance scale $R^* \sim 2/(\omega C_A L^2)$. We intentionally chose the gate capacitances, region lengths and operating frequency so that $R^*$ lies on the few-kΩ scale, placing the maximum sensitivity in the resistance range surrounding the MIT crossover. Although the sensitivity decreases away from $R^*$, the resonance frequency continues to vary monotonically with $R$ over a much broader interval, enabling resistance extraction deep into both the metallic and insulating regimes.

To convert this resonance shift into $R$ robustly against slow temperature drifts of the circuit components, we form the dimensionless quantity $\chi = (1/f_0^2 - 1/f_{\mathrm{off}}^2)/(1/f_{\mathrm{on}}^2 - 1/f_{\mathrm{off}}^2)$; $\chi$ cancels temperature drifts of the matching-network parameters at the percent level. Inverting the measured $\chi$ through the modeled $\chi(R)$ yields the channel sheet resistance $R$ [(*36*), section 3]. Figure 1F shows the resulting $\chi$ and extracted $R$ versus $V_{\mathrm{BG}}$ at $T = 2$ K, spanning $R \approx 0.5$ kΩ in the metal to ≈ 1 MΩ in the insulator.

## Density-tuned metal–insulator transition

Using the contact-free readout of charge resistivity, we map the MIT of the bilayer $MoSe_2$ channel as a function of carrier density. Figure 2 shows the extracted channel sheet resistance $R(T)$ from 2 to 20 K over a range of densities, for electron doping (Fig. 2A) and hole doping (Fig. 2D), at zero magnetic field. At low density the curves are insulating — $R$ rises by orders of magnitude on cooling — whereas at high density they are metallic, with $R$ decreasing on cooling. Between these limits the sign of d$R$/d$T$ inverts, marking the density-tuned MIT.

We estimate the critical density from the temperature dependence, using two complementary criteria(*37*), and denote the result as $n_c$. In the insulating regime the resistance is thermally activated, $R \propto \exp[E_a/(k_B T)]$; fitting over 3–6 K gives an activation energy $E_a$ that decreases with density and extrapolates to zero (blue, Fig. 2, C and F). Independently, the logarithmic slope $\alpha = \mathrm{d}\ln R/\mathrm{d}\ln T$ over 2–4 K changes sign as the channel turns metallic (red, Fig. 2, C and F). The two criteria coincide on both sides, at $n_c \approx 0.53 \times 10^{12}$ cm$^{-2}$ for electrons (Fig. 2C) and $n_c \approx 1.98 \times 10^{12}$ cm$^{-2}$ for holes (Fig. 2F). At the respective critical densities, the band masses and degeneracies introduced above give Fermi temperature $T_F = E_F/k_B \approx 5$ K for electrons and ≈ 44 K for holes. The electrons are therefore only marginally degenerate at the base temperature ($T = 2$ K), whereas the holes are deep in the degenerate regime. Here and below, the quoted $T_F$ values refer to single-particle estimates based on the corresponding band structure; interactions may modify the relevant Fermi-energy scale.

Near the metal–insulator crossover the temperature dependence becomes non-monotonic for both carriers. For electrons, non-monotonic behavior appears on the metallic side, very close to $n_c$ (arrow, Fig. 2B). From the base temperature, the resistance first rises, passes through a local maximum at ≈ 4.5 K, comparable to $T_F \approx 5$ K, and then decreases upon further warming. Similar resistance maxima near $T_F$ have been observed on the metallic side of the transition in other dilute two-dimensional systems, where they have been discussed as a Pomeranchuk-like response(*1*). On the hole side, the resistance of the curves near $n_c$ first rises to a weak local maximum at ≈ 3.5 K, falls to a minimum at ≈ 5.5 K, and then rises steadily up to 20 K (arrows, Fig. 2E). These resistance extrema temperatures are far below $T_F \approx 44$ K for holes. Near the crossover, however, several effects may be superposed in $R(T)$ and we do not assign the individual extrema to a unique microscopic process.

## An interaction-anchored ambipolar transition

The two temperature-dependence criteria of Fig. 2 identify $n_c \approx 0.53 \times 10^{12}$ cm$^{-2}$ for electrons and $\approx 1.98 \times 10^{12}$ cm$^{-2}$ for holes, differing by about a factor of four. Because the two carrier polarities are measured in the same channel, this ordering provides an internal test of interaction- and disorder-based transition criteria under the same disorder landscape. Despite their different critical densities, the two transitions occur at comparable Wigner–Seitz radii $r_s$. Using an effective dielectric constant $\epsilon_{eff} = 4.4\epsilon_0$, appropriate for a $MoSe_2$ channel encapsulated between hBN layers(*25, 38*), we obtain $r_s \approx 18.0$ on the electron side and $r_s \approx 21.7$ on the hole side. The near-coincidence follows from $r_s \propto m^*/\sqrt{n}$ — the heavier hole mass compensates for the higher hole $n_c$.

By contrast, the IRM criterion for a disorder-driven strong-localization crossover predicts the opposite trend. The condition $k_F l \sim 1$ marks the point at which the disorder-limited mean free path becomes comparable to $k_F^{-1}$, beyond which coherent quasiparticle transport is expected to break down(*15, 39*). For a two-dimensional parabolic band with total spin–valley degeneracy $g_d$, $k_F = (4\pi n/g_d)^{1/2}$, $l = \hbar k_F \mu/e$, giving $k_F l = 4\pi\hbar n\mu/(g_d e)$. Using $g_{d,e} = 12$ and $g_{d,h} = 2$, together with the effective mobilities at the respective measured critical densities, $\mu_e(n_{c,e}) \approx 940$ and $\mu_h(n_{c,h}) \approx 237$ cm$^2$ V$^{-1}$ s$^{-1}$ [(*36*), section 6], gives $k_F l \approx 0.34$ for electrons and $\approx 1.9$ for holes. Thus, the measured transitions do not occur at the same $k_F l$. Equivalently, imposing $k_F l = 1$ gives the nominal IRM crossover density $n^{IRM} = g_d e/(4\pi\hbar\mu)$ and the ratio $n_h^{IRM}/n_e^{IRM} \approx 0.66$ predicts a lower crossover density for holes than for electrons, opposite to the measured ordering. The same reversed ordering, $n_h^{IRM} < n_e^{IRM}$, is obtained using the maximum metallic-regime mobilities as an alternative disorder metric [(*36*), section 1]. Taken together, these results are incompatible with a simple disorder-driven strong-localization picture. The comparable $r_s$ values instead support an interaction-driven origin of the MIT transition in the bilayer $MoSe_2$, with the critical-density scale anchored by the Coulomb-to-kinetic-energy ratio.

## Magnetotransport near the metal–insulator transition

We further investigate the magnetotransport in fields up to $B = 11.8$ T, applied both out-of-plane ($B_\perp$) and in-plane ($B_\parallel$). Figure 3, A to D, maps the MR ratio $R(11.8\text{ T})/R(0)$, with $R$ the sheet resistance, across the $(n, T)$ plane for all four combinations of carrier and field orientation: electron $B_\perp$ (Fig. 3A) and $B_\parallel$ (Fig. 3B), and hole $B_\perp$ (Fig. 3C) and $B_\parallel$ (Fig. 3D). For every carrier and orientation, the MR ratio forms a pronounced peak as a function of density near the zero-field metal–insulator crossover (density cuts at $T = 2$ K, Fig. 3, E and G), and the peak is strongest at the lowest temperature (temperature cuts, Fig. 3, F and H). We denote the peak position as $n_{MR}$. We find $n_{MR} \approx 0.53 \times 10^{12}$ cm$^{-2}$ ($B_\perp$) and $0.51 \times 10^{12}$ cm$^{-2}$ ($B_\parallel$) for electrons, and $2.07 \times 10^{12}$ cm$^{-2}$ ($B_\perp$) and $2.03 \times 10^{12}$ cm$^{-2}$ ($B_\parallel$) for holes. The $B_\perp$ and $B_\parallel$ datasets were taken in separate cooldowns, and $n_{MR}$ agrees closely between the two field orientations for both carriers. Magneto-orbital coupling is strongly suppressed for $B_\parallel$ in the atomically thin channel, so the $B_\parallel$ peak reflects a predominantly spin-coupled response. Surprisingly, the occurrence of the $B_\perp$ peak at nearly the same density as the $B_\parallel$ peak shows that orbital coupling is not required for the appearance of the MR peak. For both carriers $n_{MR}$ essentially coincides with the zero-field critical density determined from the temperature dependence ($n_c \approx$ 0.53 and $\approx 1.98 \times 10^{12}$ cm$^{-2}$ for electrons and holes, respectively). This correspondence, together with the representative field evolutions in which the peak position remains fixed near $n_c$ as its magnitude grows with field [(*36*), section 8], ties the MR peak to the physics of the zero-field transition.

The field-orientation dependence differs sharply between the two carrier types. On the hole side the MR is nearly isotropic, with peak ratios of $\approx 5.6$ for $B_\perp$ and $\approx 5.0$ for $B_\parallel$ at 2 K (Fig. 3, G and H). On the electron side it is strongly anisotropic — the peak ratio reaches $\approx 12$ for $B_\perp$ but only $\approx 1.6$ for $B_\parallel$ at 2 K (Fig. 3, E and F). This contrast is consistent with the band character of 2L-$MoSe_2$. Holes at Γ carry no valley-contrasting Ising spin–orbit field, giving an approximately isotropic Zeeman response(*40*). By contrast, the low-density electrons occupy the Q valleys, where spin–orbit coupling produces a predominantly out-of-plane spin quantization axis(*40, 41*), and $B_\perp$ therefore couples more efficiently to the electron spins than $B_\parallel$. This anisotropic spin coupling provides the primary explanation for the electron-side field anisotropy, although orbital coupling may additionally enhance the $B_\perp$ response. Similar ambipolar magnetoresponses are also observed in a second device [(*36*), section 7]. Such magnetic-field-enhanced insulating behavior has also been observed close to MIT transitions in Si-MOSFETs and GaAs heterostructures previously(*1, 13, 42-44*).

**Critical-density magnetoresistance and a Wigner-solid/correlated-liquid interpretation**

To understand the origin of the MR enhancement near $n_c$, we examine its detailed $B_\parallel$ dependence on the hole side, where the carriers are deeply degenerate and the field response is well resolved. Figure 4A shows the sheet resistance $R(B)$ at $T = 2.2$ K for a series of hole densities. Dots are the measured data, and solid lines are logistic fits [(*36*), section 1]. Figure 4B maps the normalized differential response $S = [1/R(0)]\, dR/dB$ in the ($B$, $n$) plane, calculated from the fitted curves. The strongest enhancement of $S$ remains centered on $n_c$ and decreases on both the metallic and insulating sides. We refer to this sharply density-localized feature as the critical-density MR ($n_c$-MR) response. A substantially weaker response can also be observed at higher densities.

We can understand the $n_c$-MR as the magnetic-field counterpart of the Pomeranchuk effect in the mixed Wigner-solid and correlated-liquid regions(*9, 10*). In the regime $J \ll k_B T \ll E_F$, where $J$ is the spin-exchange energy of the Wigner solid, the transport behavior of the mixture admits a particularly simple description(*45, 46*). At $T = 2.2$ K, the very low spin-exchange scales of dilute Wigner solids(*47*) support $J \ll k_B T$, while the single-particle band estimate gives $T \ll T_F \approx 44$ K for the hole liquid. Under these conditions, the localized spins of the Wigner-solid have a Curie-like susceptibility that exceeds the Pauli-like susceptibility of the degenerate liquid. An in-plane field therefore lowers the magnetic free energy of the solid more than that of the liquid and increases the Wigner-solid fraction $x_W$(*45, 46*). Near $n_c$, the conducting liquid forms a percolation network close to losing connectivity. Here, $dR/dB = (dR/dx_W)(dx_W/dB)$, and $dR/dx_W$ can become large and even a small increase in $x_W$ can therefore disrupt conducting paths and produce a large resistance change.

Recent STM studies of hole- and electron-doped bilayer $MoSe_2$, the same material system studied here, directly resolved disorder-pinned Wigner-solid and correlated-liquid regions, with the liquid regions forming a percolating network as density increased(*29, 31*). The liquid-solid percolation thresholds for both electrons and holes were observed at $r_s$ around 20(*29, 31*). (The STM and transport measurements involve different dielectric environments — the open surface TMD in STM has a lower dielectric constant than the symmetric hBN encapsulation in the transport device.) The real-space observation of solid–liquid coexistence and percolation therefore provides a microscopic basis for our transport interpretation.

Both transitions in our ambipolar transport measurements occur at $r_s$ values below the clean-limit crystallization threshold $r_s \gtrsim 30$(*4, 6-8*). Quenched disorder can lower this threshold: it pins the Wigner solid and stabilizes solid–liquid coexistence(*48*), and calculations including impurities place the transition at substantially lower $r_s$ than in the clean system(*49*).

## Conclusion

Using capacitively coupled radio-frequency reflectometry, we have measured the ambipolar metal–insulator transition in a single 2L-$MoSe_2$ channel, including the hole-side transition previously inaccessible to conventional transport. The $n_c$ values for electrons and holes differ by a factor of four yet occur at comparable Coulomb-to-kinetic-energy ratios, quantified by Wigner–Seitz radii $r_s \approx 18$ (electron) and 22 (hole). Magnetotransport further reveals a pronounced MR sharply localized near the $n_c$ of each carrier type. Together with the Wigner-solid and correlated-liquid coexistence resolved by STM in this material system, we interpret the critical-density MR response as arising from a field-induced change in the balance between them, amplified by the marginal connectivity of the liquid network near $n_c$. Our results thus address two central questions surrounding the two-dimensional MIT, linking the zero-field transition scale to interaction strength in relatively clean 2D semiconductors, and the MR anomaly to microscopic solid–liquid coexistence. More broadly, the resistance readout relies on capacitive gate coupling and does not require low-resistance ohmic contacts to the target material. It therefore enables quantitative transport measurements in a broad class of gate-tunable two-dimensional systems that have so far been inaccessible because of large contact resistance.

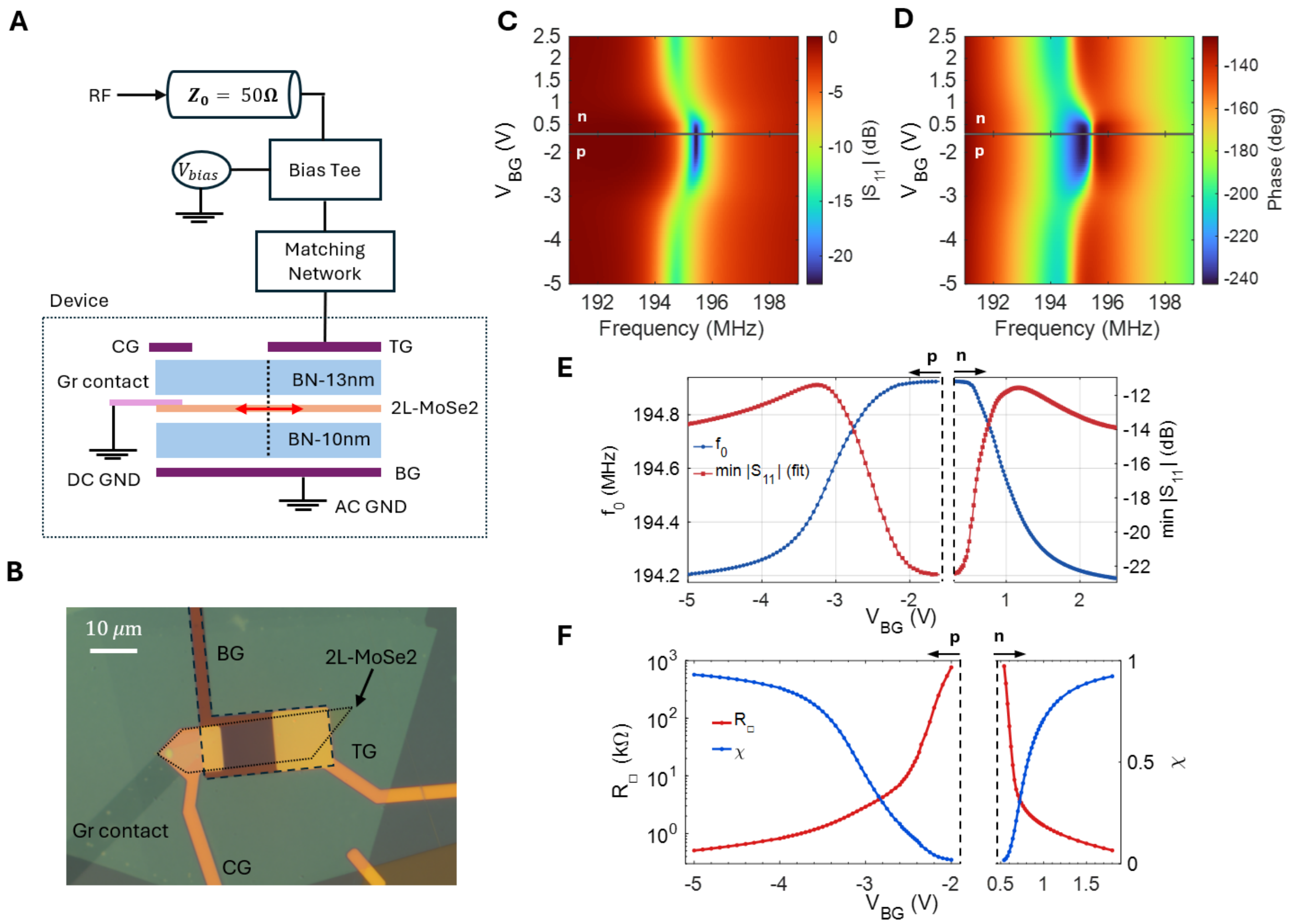

**Fig. 1. Radio-frequency reflectometry of bilayer $MoSe_2$.** (**A**) Circuit schematic and device cross-section. The RF signal is applied to the top-gate (TG) through a bias tee and an impedance-matching network, while the back-gate (BG) is AC-grounded. The bilayer $MoSe_2$ (2L-$MoSe_2$) channel is encapsulated between 13-nm top and 10-nm bottom hBN dielectrics and connected to a graphite contact (Gr contact) held at DC ground. The contact-gate (CG) dopes the contact region to maintain rapid carrier equilibration during gate sweeps. At the RF measurement frequency this contact is effectively open; the readout therefore probes the in-plane motion of carriers through capacitive coupling to the gates. The red arrow indicates the in-plane oscillation of carriers driven by the RF excitation; the dotted line marks the boundary between the dual-gated and back-gate-only portions of the channel. (**B**) Optical micrograph of the device (scale bar, 10 μm). Dashed lines outline the BG and the 2L-$MoSe_2$ flake. (**C** and **D**) Reflected amplitude $|S_{11}|$ (**C**) and phase (**D**) spectra as a function of back-gate voltage $V_{BG}$ at $T = 2$ K. The n-doped (upper) and p-doped (lower) regimes were measured separately; $V_{TG}$ is set to keep the two regions density matched throughout the sweep. The charge-neutral region is omitted. (**E**) Resonance frequency $f_0$ (blue, left axis) and minimum $|S_{11}|$ (red, right axis) versus $V_{BG}$, extracted from the global line-shape fit. The charge-neutral region is omitted (broken axis); vertical dashed lines mark the onset of conduction. (**F**) Extracted channel sheet resistance $R$ (red, left axis) and the dimensionless inversion fraction $\chi$ (blue, right axis) versus $V_{BG}$ at $T = 2$ K; $R$ is obtained by inverting the measured $\chi$ through the modeled $\chi(R)$.

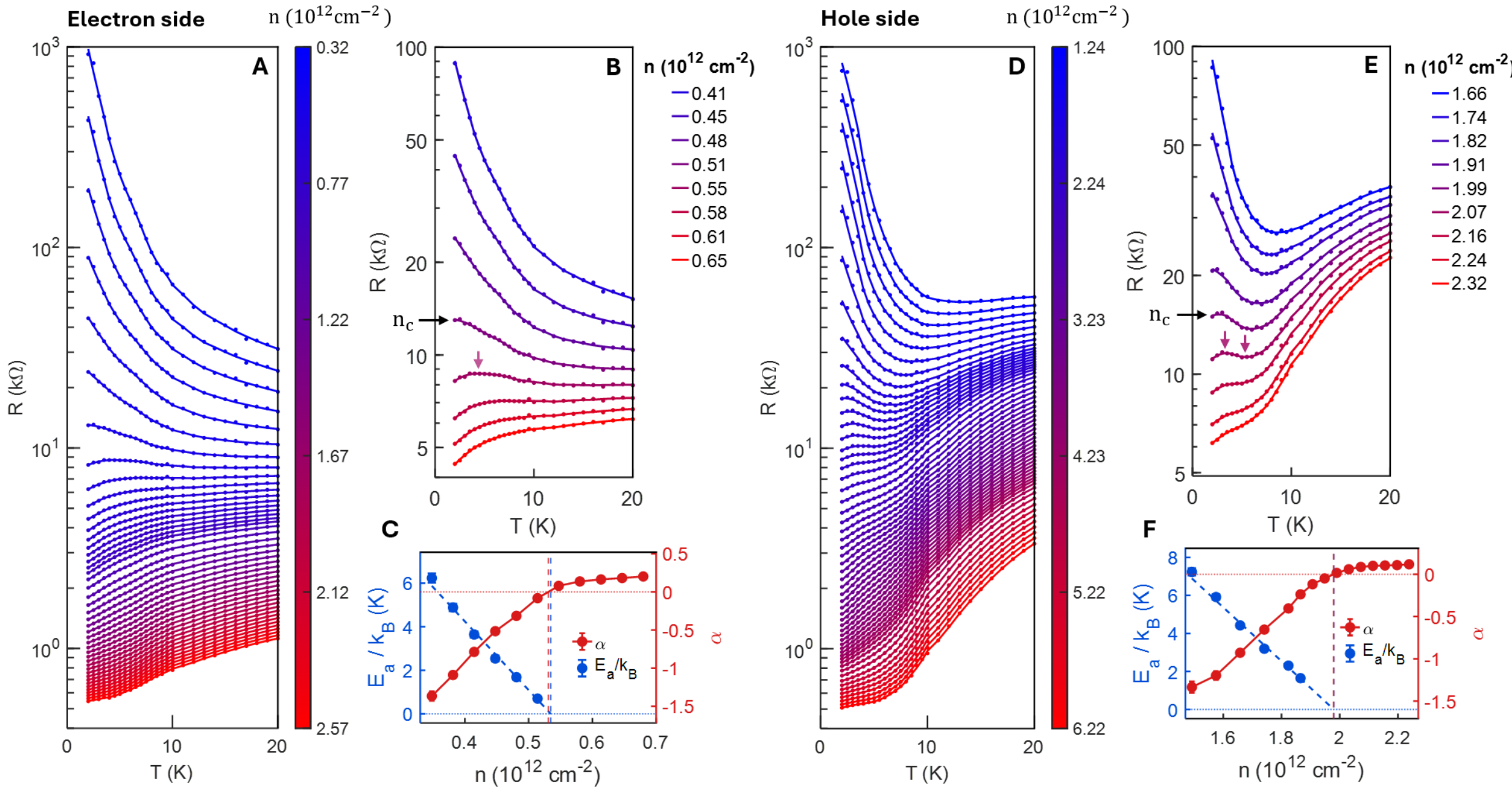

**Fig. 2. Density-tuned metal–insulator transition of bilayer $MoSe_2$.** (**A**) Channel sheet resistance $R$ versus temperature $T$ (2–20 K) on the electron side, at carrier densities $n = 0.32–2.57 \times 10^{12}$ cm$^{-2}$ (color scale). The curves evolve from insulating ($R$ increasing on cooling; blue) to metallic ($R$ decreasing on cooling; red). (**B**) Magnified view of the electron-side crossover range ($n = 0.41–0.65 \times 10^{12}$ cm$^{-2}$). The labeled arrow marks the curve nearest to $n_c$; the small arrow marks the local resistance maximum at ≈ 4.5 K on the metallic side. (**C**) Activation energy $E_a/k_B$ (blue, left axis) and the low-temperature logarithmic slope $\alpha = \mathrm{d}\ln R/\mathrm{d}\ln T$ evaluated over 2–4 K (red, right axis), versus $n$. $E_a$ is obtained by fitting $R \propto \exp[E_a/(k_B T)]$ over 3–6 K. The temperature-based critical density $n_c$ is identified where $E_a \to 0$ and where $\alpha$ changes sign; the two criteria coincide at $n_c \approx 0.53 \times 10^{12}$ cm$^{-2}$ (dashed line). (**D** to **F**) Corresponding data for the hole side [$n = 1.24–6.22 \times 10^{12}$ cm$^{-2}$ in (**D**); $1.66–2.32 \times 10^{12}$ cm$^{-2}$ in (**E**)]. In (**E**), the labeled arrow marks the curve nearest to $n_c$, and the small arrows mark the local maximum at ≈ 3.5 K and the minimum at ≈ 5.5 K of the curves near $n_c$. In (**F**), the two criteria coincide at $n_c \approx 1.98 \times 10^{12}$ cm$^{-2}$ (dashed line). All data at zero magnetic field.

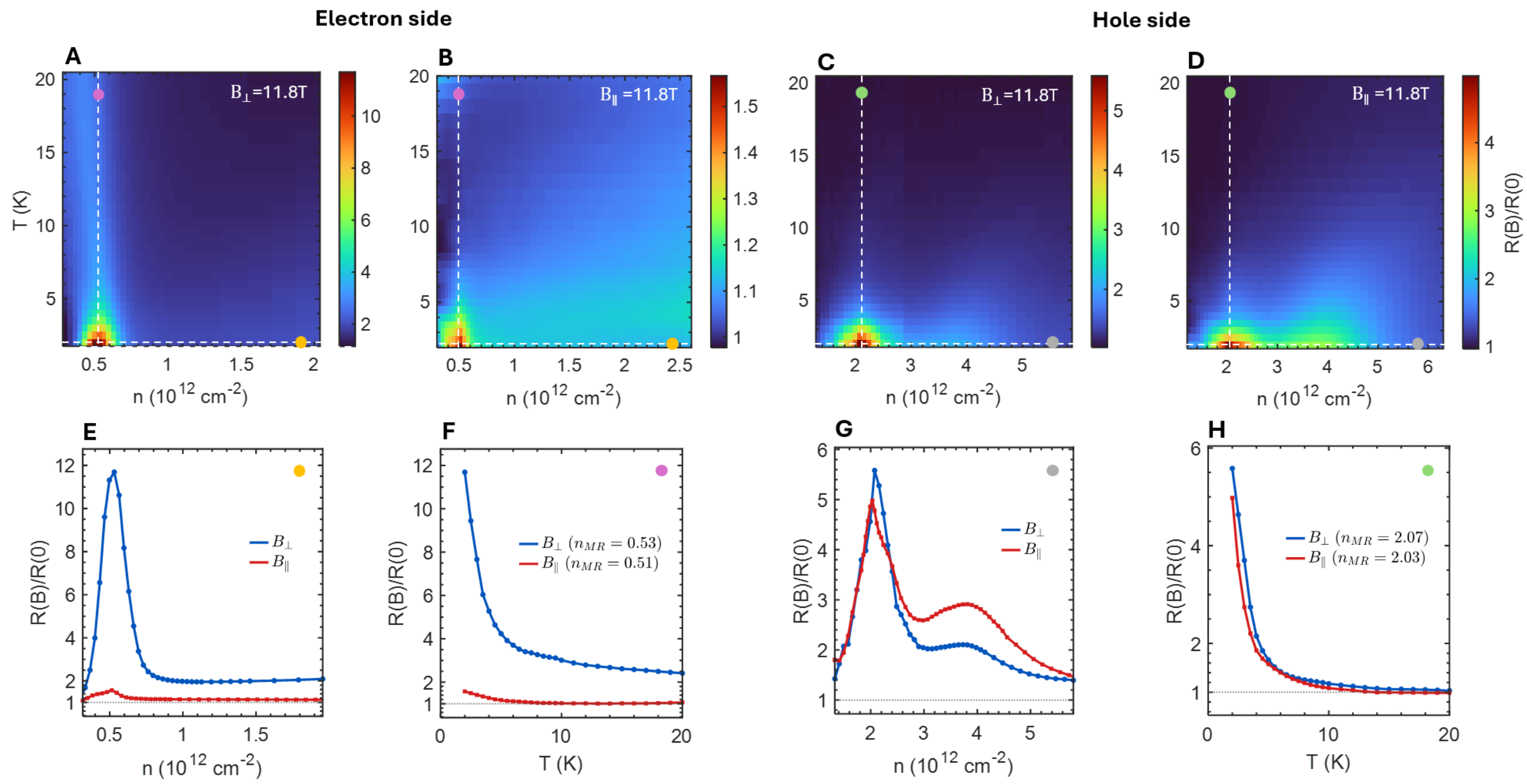

**Fig. 3. In-plane and out-of-plane magnetotransport across the metal–insulator transition.** (**A** to **D**) Magnetoresistance (MR) ratio of the sheet resistance, $R(B)/R(0)$ at $B = 11.8$ T across the ($n$, $T$) plane, for electrons in out-of-plane ($B_\perp$) (**A**) and in-plane ($B_\parallel$) (**B**) fields and for holes in $B_\perp$ (**C**) and $B_\parallel$ (**D**) fields. For every carrier and orientation, the ratio exhibits a pronounced peak at $n_{\mathrm{MR}}$, which lies close to the temperature-based critical-density estimate $n_c$ from Fig. 2. The vertical dashed line marks $n_{\mathrm{MR}}$ and the horizontal dashed line marks $T = 2$ K; together with the markers, they indicate the temperature and density cuts shown below. (**E** and **G**) MR ratio as a function of density at $T = 2$ K, with $B_\perp$ (blue) and $B_\parallel$ (red) overlaid, for electrons (**E**) and holes (**G**); the hole side shows a second, broader maximum near $n \approx 3.5$–$4.2 \times 10^{12}$ cm$^{-2}$. (**F** and **H**) MR ratio as a function of temperature at $n_{\mathrm{MR}}$, with $B_\perp$ and $B_\parallel$ overlaid, for electrons [(**F**); $n_{\mathrm{MR}} \approx 0.53 \times 10^{12}$ cm$^{-2}$ for $B_\perp$ and $0.51 \times 10^{12}$ cm$^{-2}$ for $B_\parallel$] and holes [(**H**); $n_{\mathrm{MR}} \approx 2.07 \times 10^{12}$ cm$^{-2}$ for $B_\perp$ and $2.03 \times 10^{12}$ cm$^{-2}$ for $B_\parallel$]. The two orientations were measured in separate cooldowns, and the $n_{\mathrm{MR}}$ values agree closely between them.

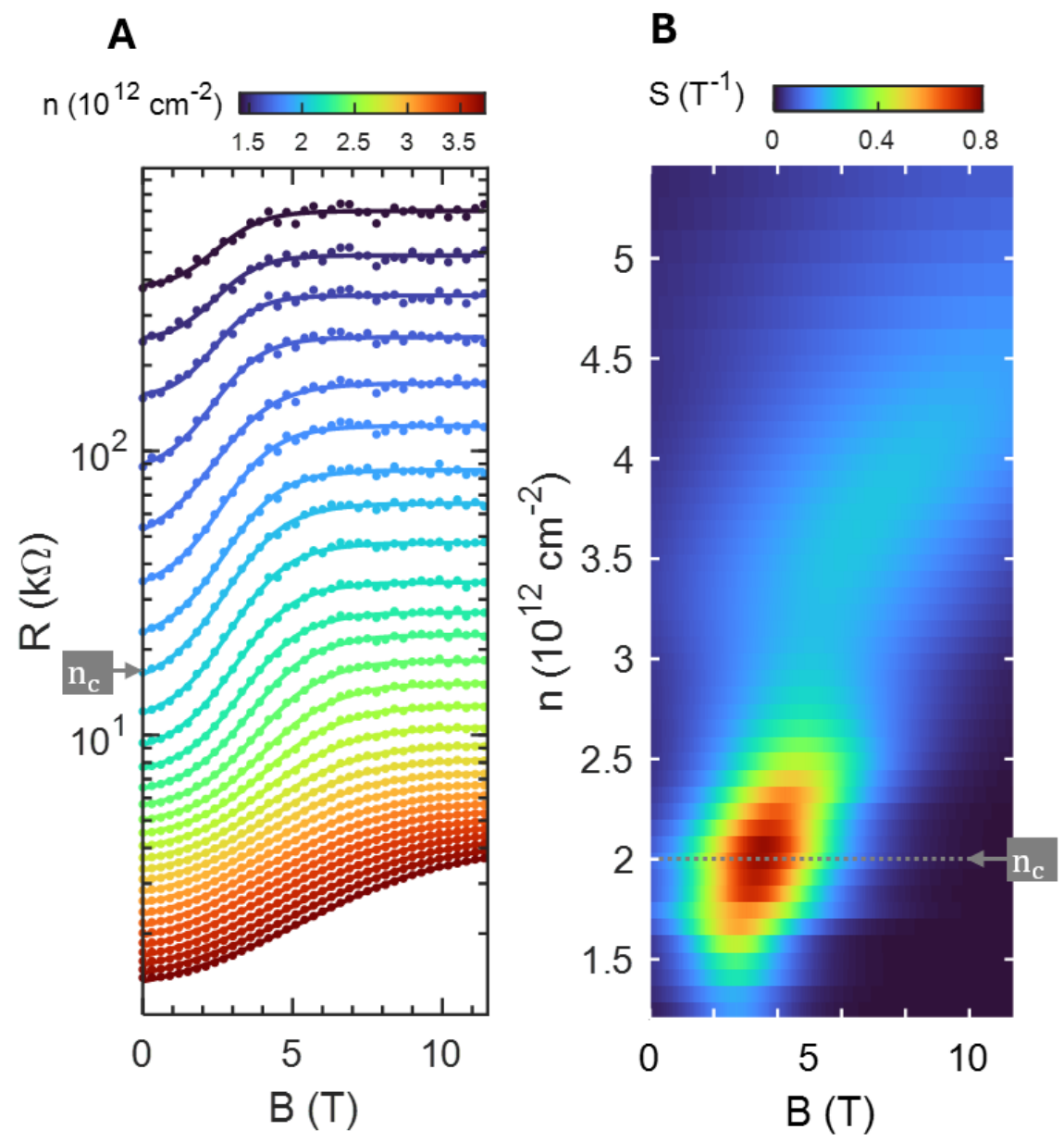

**Fig. 4. Field-resolved in-plane critical-density magnetoresistance (hole side).** (**A**) Sheet resistance $R$ versus in-plane magnetic field $B$ at $T = 2.2$ K for a series of hole densities $n$ (color scale). Dots are measured data and solid lines are logistic fits [(*36*), section 1]. The gray arrow marks the zero-field critical density $n_c$. (**B**) Normalized differential response $S = [1/R(0)]\ dR/dB$ in the $(B, n)$ plane computed from the fitted curves of (**A**). The strong enhancement of $S$ is centered at $n_c \approx 2.0 \times 10^{12}$ cm$^{-2}$ (dotted line); a much weaker response at higher densities is widely distributed over the measured field range.

## Acknowledgments

**Funding:** This work is supported by the Director, Office of Science, Office of Basic Energy Sciences, Materials Sciences and Engineering Division of the US Department of Energy under contract number DE-AC02-05CH11231 (van der Waals heterostructure program KCWF16). K.W. and T.T. acknowledge support from the CREST (JPMJCR24A5), JST and World Premier International Research Center Initiative (WPI), MEXT, Japan.

**Author contributions:** F.W. and H.-L.K. proposed and designed the experiment; H.-L.K. fabricated the devices with the help of H.L. and T.W.; H.-L.K. designed and fabricated the RF circuit boards with input from F.W., E.Y.M. and A.Y.W.; A.R. and G.G. provided access to a separate cryostat for preliminary low-temperature tests of the RF circuit; H.-L.K. built the cryogenic RF setup, including the low-temperature RF lines in the measurement cryostat, and conducted all measurements; H.-L.K. and F.W. analyzed the data. K.W. and T.T. grew hBN crystals. H.-L.K. and F.W. wrote the manuscript with input from all authors.

**Competing interests:** The authors declare that they have no competing interests.

**Data and materials availability:** All data are available in the manuscript or the supplementary materials.

**List of Supplementary Materials**: Materials and Methods; Supplementary Text; Figs. S1 to S11; Table S1; References (50–57)

## Supplementary Materials for

# Ambipolar metal–insulator transition in bilayer $MoSe_2$

Ha-Leem Kim*, Hyungbin Lim, Terry Wu, Amogh Y. Waghmare, Garima Gupta, Takashi Taniguchi, Kenji Watanabe, Archana Raja, Eric Y. Ma, Feng Wang*

*Corresponding author. Email: haleem4644@berkeley.edu (H.-L.K.); fengwang76@berkeley.edu (F.W.)

**This PDF file includes:**

Materials and Methods
Supplementary Text
Figs. S1 to S11
Table S1
References (50–57)

# Contents

## Section 1: Materials and Methods

**Device fabrication.** Bilayer $MoSe_2$ (2L-$MoSe_2$), from HQ Graphene, few-layer graphite, and hexagonal boron nitride (hBN) were mechanically exfoliated onto 90-nm $SiO_2$/Si wafers. The 2L-$MoSe_2$ channel was encapsulated between a 13-nm top and a 10-nm bottom hBN dielectric and contacted at one end by a few-layer graphite electrode. Heterostructures were assembled with polycarbonate (PC) film stamps and released onto an intrinsic (high-resistivity) Si substrate, which keeps the parasitic capacitance low. Rather than a conventional high-temperature melt, the PC film was released at 45 °C by locally dissolving the PC with a small volume of chloroform delivered through a 5-μL Hamilton microsyringe (32s-gauge needle); this low-temperature release reduces strain that otherwise accumulates during high-temperature release. The top-gate and contact-gate (Cr/Au, 3/30 nm) and the back-gate (Cr/Au, 3/12 nm) were defined by electron-beam lithography and metal evaporation, with the gold kept thin to minimize strain from the deposited metal; wire bond pads were patterned by photolithography and metal evaporation (Cr/Au, 5/120 nm). The top-gate covers approximately half of the channel, defining the dual-gated Region 2 (length $L_2 \approx$ 14 μm), while Region 1 ($L_1 \approx$ 16 μm), controlled by the back-gate alone, adjoins the graphite contact; the channel width is $W \approx$ 11 μm.

**RF reflectometry measurement.** Reflectometry was performed with a custom radio-frequency circuit mounted at the sample stage. The bias-tee and matching-network design and the choice of cryogenically compatible components were informed by the RF-reflectometry literature(*50*). An on-board bias tee combines the RF stimulus with a DC bias on the top-gate, and a C–L–C matching network (shunt $C_{match}$ = 100 pF, series air-core inductor $L_c$ = 110 nH, and an additional shunt capacitor setting the total device-side shunt capacitance to ~ 6 pF) impedance-matches the device to the 50-Ω line, forming a resonance in the 180–200 MHz range. The back-gate is AC-grounded through a 1-nF shunt capacitor. The complex reflection coefficient $S_{11}(f)$ was recorded with a vector network analyzer (Siglent SVA1015X). The sample was mounted in the variable-temperature insert of a 12-T cryostat (Cryomagnetics C-Mag Vari-12T), and data span $T$ = 2–20 K and magnetic fields up to $B$ = 11.8 T applied in-plane or out-of-plane (see section 2 for details of the cryogenic RF setup and circuit design). The RF power at the 50 Ω input port feeding the matching-network/sample assembly was calibrated to $P_{in} = -92$ dBm after accounting for room-temperature and cryogenic attenuation, cable losses and directional-coupler loss. A directional coupler inside the cryostat separated the incident and reflected waves, and the reflected signal was amplified by a chain of three room-temperature low-noise amplifiers with a total gain of 72 dB before detection. At this input power, the distributed-RC simulation gives a peak-to-peak top-gate-node voltage below 0.24 mV, a peak lateral sheet current density below $1.7\ \mathrm{nA}\, \mu m^{-1}$, and a maximum time-averaged Joule heating per unit channel length below $0.016\ \mathrm{pW}\, \mu m^{-1}$ across the modeled resistance range (section 3). No measurable power dependence was observed when the input power was reduced from −92 to −102 dBm, so we treat the readout as non-perturbative under these conditions. The top-gate DC bias was set to keep the two regions density matched throughout the back-gate sweep (section 4), and the contact-gate was biased to keep the graphite-contact region conductive for rapid carrier equilibration. Our RF readout is a device-integrated counterpart of microwave impedance microscopy, which senses local conductivity and permittivity through a capacitively coupled tip(*51, 52*). Using the top-gate as the coupling electrode reads out the entire channel through a known capacitance, which gives an absolute sheet resistance and makes dense maps versus density, temperature and magnetic field practical.

**Resistance and magnetotransport analysis.** At a given magnetic field and temperature, gate-dependent $S_{11}(f)$ was fitted with a global complex-Lorentzian line-shape model (section 5) to extract the resonance frequency $f_0$. The channel sheet resistance was obtained by calculating $\chi$ from $f_0$ and inverting the measured $\chi$ through a modeled $\chi(R)$ curve computed from the distributed-RC model embedded in the full RF matching network (section 3). Carrier density was calculated by $n = \epsilon_r\epsilon_0|V_{BG} - V_0|/de$, where $\epsilon_0$ is the vacuum permittivity, $\epsilon_r = 3.0$(*53*) is the DC dielectric constant of hBN along the out-of-plane direction, d is the thickness of the back-gate hBN dielectric measured by AFM and $V_0$ is the carrier-onset voltage for the corresponding polarity. Mobilities were obtained as $\mu = 1/(enR)$ (Fig. S6). At $T = 2.0$ K, the mobilities at the respective measured critical densities are $\mu_e(n_{c,e}) \approx 940$ cm$^2$ V$^{-1}$ s$^{-1}$ at $n_{c,e} \approx 0.53 \times 10^{12}$ cm$^{-2}$ and $\mu_h(n_{c,h}) \approx 237$ cm$^2$ V$^{-1}$ s$^{-1}$ at $n_{c,h} \approx 1.98 \times 10^{12}$ cm$^{-2}$. The maximum mobilities in the high-density metallic regime are $\mu_e^{max} \approx 4.44 \times 10^3$ cm$^2$ V$^{-1}$ s$^{-1}$ and $\mu_h^{max} \approx 1.97 \times 10^3$ cm$^2$ V$^{-1}$ s$^{-1}$, which we use as alternative disorder metrics in the Ioffe–Regel–Mott (IRM) comparison. For the IRM analysis, $k_F = (4\pi n/g_d)^{1/2}$ and $l = \hbar k_F\mu/e$, giving $k_F l = 4\pi\hbar n\mu/(g_d e)$. We use $g_{d,e} = 12$ for the six spin-degenerate Q valleys and $g_{d,h} = 2$ for the spin-degenerate Γ hole band. Evaluating this expression at the measured critical densities gives $k_F l \approx 0.34$ for electrons and $\approx 1.9$ for holes. Imposing $k_F l = 1$ defines the nominal IRM crossover density $n^{IRM} = g_d e/(4\pi\hbar\mu)$. Inserting the mobilities at the measured critical densities gives $n_e^{IRM} \approx 1.54 \times 10^{12}$ cm$^{-2}$ and $n_h^{IRM} \approx 1.02 \times 10^{12}$ cm$^{-2}$, corresponding to $n_h^{IRM}/n_e^{IRM} \approx 0.66$. Using the maximum metallic-regime mobilities instead gives $n_h^{IRM}/n_e^{IRM} \approx 0.38$. The Wigner–Seitz radius $r_s = 1/(\sqrt{(\pi n)}\, a_B^*)$ was calculated using $a_B^* = \epsilon_{eff}(m_e/m^*)a_0$, with $\epsilon_{eff} = 4.4$(*25, 38*), $m^* = 0.54\, m_e$ for the Q-valley electron band and $m^* = 1.26\, m_e$ for the Γ-valley hole band, using bilayer $MoSe_2$ band-structure parameters(*29*). For the magnetotransport analysis, each $R(B)$ trace was fitted to a logistic form $R(B) = R_{low} + \Delta R/[1 + \exp(-(B - B_0)/w)]$. The normalized differential response $S(B, n) = [1/R(0)]\, dR/dB = d\mathrm{MR}/dB$, where $\mathrm{MR} = [R(B) - R(0)]/R(0)$, was computed from the fitted curves.

## Section 2: Cryogenic RF measurement setup and circuit design

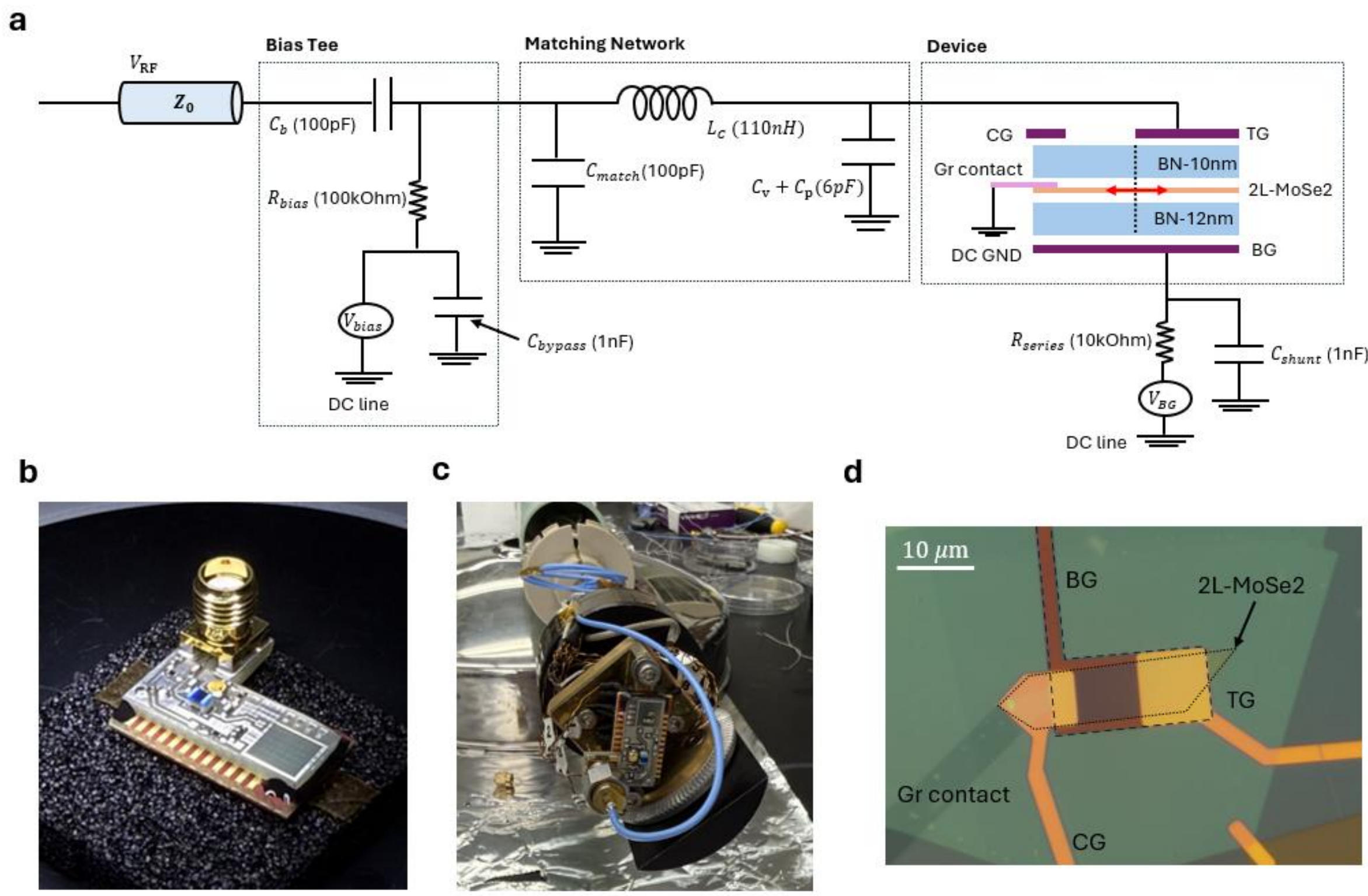

**Fig. S1. Cryogenic radio-frequency (RF) reflectometry setup for S11 measurements**. **a**, Circuit schematic detailing the custom bias tee, matching network, and 2L-$MoSe_2$ device connections. **b**, Photograph of the custom-fabricated RF printed circuit board (PCB) mounted on a 16-pin adapter. **c**, Photograph of the PCB and wiring installed within the variable temperature insert cryostat sample space. **d**, Optical micrograph of the fabricated 2L-$MoSe_2$ device prior to RF integration, highlighting the top and back gate geometries.

### 1. Circuit Design and Components

To perform high-frequency reflectometry on our bilayer $MoSe_2$ device, we used a custom-designed RF circuit mounted directly at the sample stage (Fig. S1a,b,c). The bias-tee and matching-network design and the choice of cryogenically compatible components were informed by the RF-reflectometry literature(*50, 54, 55*). The input line with a characteristic impedance $Z_0$ = 50 Ω enters an on-board bias tee. This bias tee consists of a blocking capacitor $C_{block}$ (100 pF, Kemet CBR04C101F3GAC), a bias resistor $R_{bias}$ (100 kΩ, Digi-Key P100KDCDKR-ND), and a bypass capacitor $C_{bypass}$ (1 nF, Kemet C0402C102J5GECAUTO), allowing for the simultaneous application of an RF stimulus and a DC bias to the top gate.

The signal then passes through a C–L–C matching network designed to impedance-match the device to the 50 Ω line. The network uses a shunt capacitor $C_{match}$ (100 pF) and a series air-core inductor $L_c$ (110 nH, Murata LQW2BASR11J00L). An additional shunt capacitor $C_v$ was attached to the device side so that the total shunt capacitance at the device side is ≈ 6 pF, accounting for the intrinsic parasitic capacitance ($C_p$) of the device. Using an intrinsic Si substrate reduces the parasitic capacitance to below approximately 2 pF.

The device itself is a 2L-$MoSe_2$ channel encapsulated between 13 nm top and 10 nm bottom hexagonal boron nitride (hBN) dielectrics (Fig. S1d). The top gate (TG), consisting of deposited Cr/Au (3 nm / 30 nm), covers only half of the 2L-$MoSe_2$ region. The back gate (BG) is composed of Cr/Au (3 nm / 12 nm). The BG is controlled via a separate DC line ($V_{BG}$) filtered by a 10 kΩ series resistor $R_{series}$ (Digi-Key P10KDCDKR-ND) and a 1 nF shunt capacitor $C_{shunt}$. This 1 nF shunt capacitor ensures the BG acts as an effective AC ground.

## 2. Custom PCB Fabrication

To minimize parasitic losses and tightly integrate the bias tee and matching network with the sample, we fabricated a custom PCB (Fig. S1b). The board was patterned on single-sided copper-coated FR4 (MG Chemicals, Part No. 503) using standard photolithography and wet ferric chloride etching techniques. This was followed by a thin tin protective coating over the remaining copper traces, which increases the shelf life of the PCB.

The RF input is interfaced via a vertical-mount SMA connector (Pasternack PE4910). To securely mount the PCB inside the cryostat, it is mechanically and electrically attached to a 16-pin DIP adapter (Aries Electronics 16-350000-11-RC), which plugs directly into the cryostat's standard 16-pin sample socket. Mechanical connections were made using Stycast 2850FT, and electrical connections between the 16-pin adapter and the PCB were made via thick Cu wire.

## 3. Cryogenic Integration and Signal Routing

The entire sample assembly is housed within a Cryomagnetics, Inc. C-Mag Vari-12T research cryostat (Fig. S1c). Reflectometry measurements are driven and recorded using a vector network analyzer (Siglent SVA1015X).

The RF power at the 50 Ω input port feeding the matching-network/sample assembly was calibrated to $P_{in} = -92$ dBm after accounting for room-temperature attenuation (Mini-Circuits BW-S20W2+), cryogenic attenuation (Amphenol XMA 2082-6043-20-CRYO), cable losses and directional-coupler loss. The primary RF lines running down the cryostat probe from room temperature to low temperature are custom-wired assemblies (Suzhou Talent Microwave Inc. G2SS-SMAMSMAM-1M).

A directional coupler (Mini-Circuits ZX30-12-4-S+) is located inside the cryostat and separates the incident and reflected waves. The principal instruments and circuit components are listed in Table S1. The local connection from this directional coupler to the sample PCB is made using a highly flexible 304SS MicroFlex Coax Cable (Quantum Coax). The reflected signal from the device is then routed back to the VNA after passing through a coupler and a series of three room-temperature low-noise amplifiers (Mini-Circuits ZX60-33LN-S+) providing a total gain of 72 dB to maximize the signal-to-noise ratio before detection.

**Table S1. Components used in the cryogenic RF circuit.**

| Component Category | Description / Value | MFR / Supplier | Part Number / Model |
|---|---|---|---|
| **Instruments & Core Systems** | | | |
| Cryostat | VTI Research System | Cryomagnetics, Inc. | C-Mag Vari-12T |
| VNA | Vector network analyzer | Siglent | SVA1015X |
| **RF Components** | | | |
| Directional Coupler | Separate incident/reflected waves | Mini-Circuits | ZX30-12-4-S+ |
| Low-Noise Amplifier | Three amplifiers used for signal amplification | Mini-Circuits | ZX60-33LN-S+ |
| Attenuator (Room Temp) | 20 dB attenuation | Mini-Circuits | BW-S20W2+ |
| Attenuator (Cryogenic) | 20 dB cryogenic attenuation | Amphenol XMA | 2082-6043-20-CRYO |
| **Cabling & Connectors** | | | |
| Coaxial Cable (Primary) | Custom-wired RF assembly | Suzhou Talent Microwave | G2SS-SMAMSMAM-1M |
| Coaxial Cable (Local) | Highly flexible 304SS MicroFlex | Quantum Coax | 304SS MicroFlex |
| SMA Connector | Vertical mount for PCB interfacing | Pasternack | PE4910 |
| **Custom PCB & Mounting** | | | |
| PCB Material | Single-sided copper-coated FR4 | MG Chemicals | 503 |
| DIP Adapter | 16-pin adapter for cryostat socket | Aries Electronics | 16-350000-11-RC |
| **Passive Circuit Components** | | | |
| Inductor ($L_c$) | 110 nH, air core | Murata | LQW2BASR11J00L |
| Capacitor ($C_{block}$, $C_{match}$) | 100 pF | Kemet | CBR04C101F3GAC |
| Capacitor ($C_{bypass}$, $C_{shunt}$) | 1 nF | Kemet | C0402C102J5GECAUTO |
| Capacitor ($C_v$) | ≈ 5 pF | Johanson Tech Inc. | S402DS |
| Resistor ($R_{series}$) | 10 kΩ | Panasonic Electronic Component | P10KDCDKR-ND |
| Resistor ($R_{bias}$) | 100 kΩ | Panasonic Electronic Component | P100KDCDKR-ND |

## Section 3: Distributed-RC model and χ-inversion extraction of the channel resistance

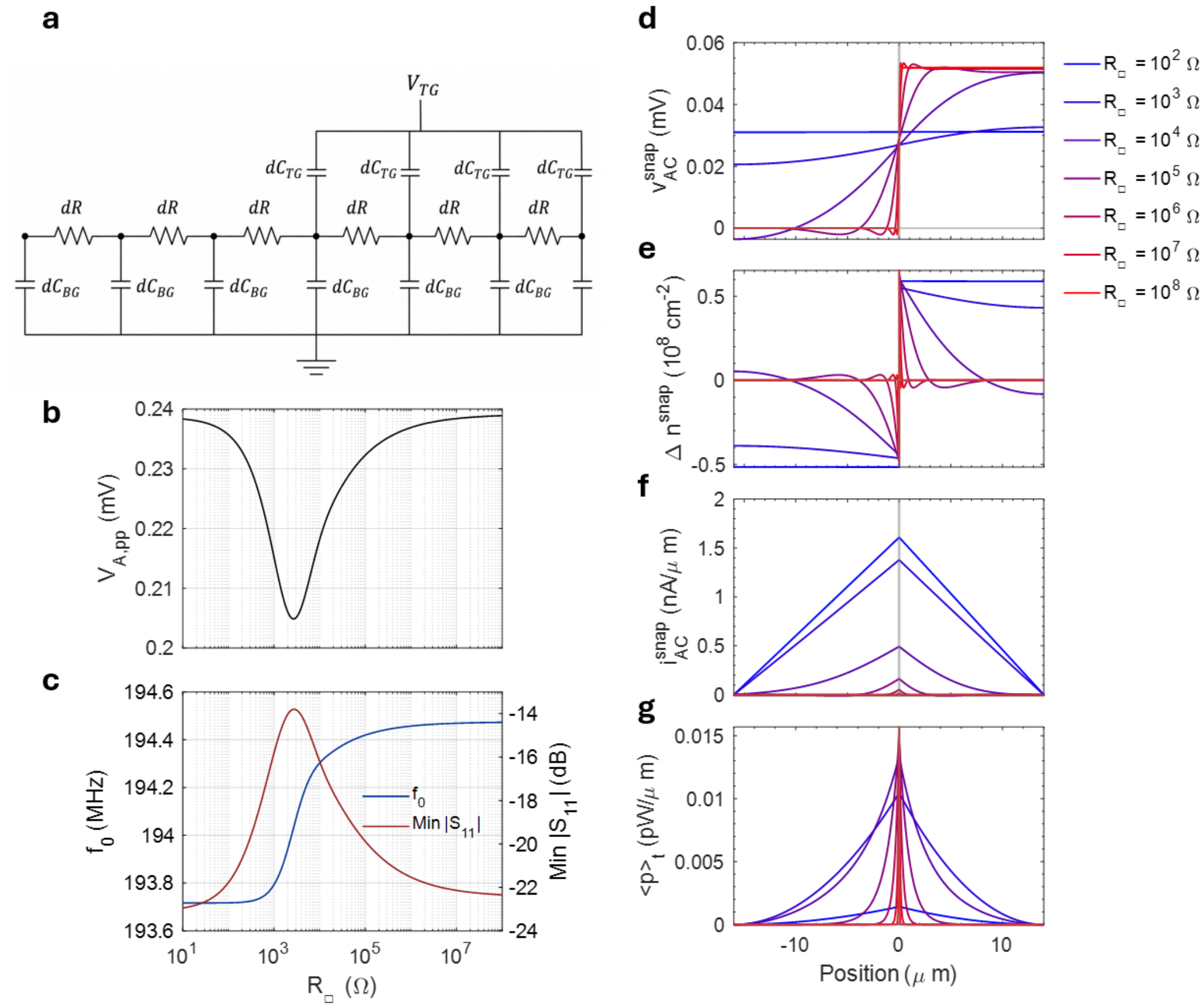

**Fig. S2. Distributed-RC model**. **a**, Equivalent-circuit representation of the two-region channel. A differential segment of length $dx$ contributes a series resistance $dR = r\,dx$, where $r = R/W$. Region 1 is coupled only to the AC-grounded back gate through $dC_{BG} = c_{BG}dx$. Region 2 is coupled to the top-gate node and to the AC-grounded back gate through $dC_{TG} = c_{TG}dx$ and $dC_{BG}$, respectively. Both channel ends are treated as RF open boundaries. The discrete segments are schematic; the calculation uses the continuous limit. **b**, Simulated peak-to-peak voltage $V_{A,pp}$ at the top-gate node, evaluated at resonance frequency $f_0(R)$. **c**, Simulated $f_0$ and minimum reflected amplitude 20 $\log_{10}|S_{11}|$ versus $R$. **d–f**, Peak-phase spatial profiles of the channel voltage $v_{AC}^{snap}(x)$, carrier-density modulation $\Delta n^{snap}(x)$ and lateral sheet current density $i_{AC}^{snap}(x)$. For each quantity and each $R$, the RF phase is chosen independently so that the largest absolute value of that quantity appears as a positive maximum. These panels are therefore not simultaneous snapshots of a single RF cycle. **g**, Time-averaged Joule heating per unit channel length, $\langle p \rangle_t$. The profiles in **d–g** are evaluated at the corresponding $f_0(R)$. Position $x = 0$ marks the boundary between Region 1 and Region 2. The calculation uses $W = 11$ μm, $L_1 = 16$ μm, $L_2 = 14$ μm, $t_{BG} = 10$ nm, $t_{TG} = 13$ nm, $C_{sh} = C_p + C_v = 6.285$ pF, $L_c = 110$ nH, inductor quality factor $Q_L$ = 95.5, $C_{match}$ = $C_{block}$ = 100 pF and $P_{in}$ = −92 dBm at the 50 Ω input port feeding the matching-network/sample assembly.

**Distributed-RC description**

At the measurement frequency, both the channel resistance and the gate capacitance are distributed along the bilayer $MoSe_2$ channel. We therefore represent the channel by two coupled distributed-RC sections and embed their calculated top-gate admittance in the complete matching network described in section 2. The model assigns the same effective sheet resistance $R$ to Region 1 and Region 2, consistent with the density-matched gate configuration used in the measurements.

**Geometry and capacitance parameters**

Region 1 has length $L_1$ = 16 $\mu$m and is coupled only to the back gate. Region 2 has length $L_2$ = 14 $\mu$m and is coupled to both the top and back gates. The channel width is $W$ = 11 $\mu$m. The top- and bottom-hBN thicknesses are $t_{TG}$ = 13 nm and $t_{BG}$ = 10 nm, respectively. Using $\epsilon_{hBN}$ = 3, the capacitances per unit area are

$$C_{TG,A} = \frac{\epsilon_0 \epsilon_{hBN}}{t_{TG}}, \qquad C_{BG,A} = \frac{\epsilon_0 \epsilon_{hBN}}{t_{BG}}.$$

The capacitances per unit channel length are

$$c_{TG} = C_{TG,A} W, \qquad c_{BG} = C_{BG,A} W,$$

and the series resistance per unit length is

$$r = \frac{R}{W}.$$

For Region 2 we define

$$c_{tot} = c_{TG} + c_{BG}, \qquad \alpha = \frac{c_{TG}}{c_{tot}}, \qquad \beta = \frac{c_{BG}}{c_{tot}} = 1 - \alpha.$$

The resistance-independent device-side shunt capacitance is

$$C_{sh} \equiv C_p + C_v = 6.285\, pF.$$

Here $C_p$ denotes the device parasitic capacitance (bonding wire + capacitance from Si substrate to wire bonding pad) excluding gates geometric capacitance and $C_v$ is the deliberately added shunt capacitor.

**Distributed-line equations and boundary conditions**

We use phasors with time dependence $e^{i\omega t}$. In Region 1,

$$\frac{dV_1}{dx} = rI_1, \qquad \frac{dI_1}{dx} = i\omega c_{\mathrm{BG}} V_1,$$

so

$$\frac{d^2V_1}{dx^2} = \gamma_1^2 V_1, \qquad \gamma_1 = \sqrt{i\omega r c_{\mathrm{BG}}}, \qquad Z_{c,1} = \sqrt{\frac{r}{i\omega c_{\mathrm{BG}}}}.$$

In Region 2, the top-gate node is driven at voltage $V_{\mathrm{A}}$. The distributed gate charge gives

$$\frac{dV_2}{dx} = rI_2,$$

$$\frac{dI_2}{dx} = i\omega[c_{\mathrm{BG}} V_2 + c_{\mathrm{TG}}(V_2 - V_A)] = i\omega c_{\mathrm{tot}}(V_2 - \alpha V_A).$$

Thus the shifted voltage $U_2 = V_2 - \alpha V_{\mathrm{A}}$ obeys

$$\frac{d^2U_2}{dx^2} = \gamma_2^2 U_2, \qquad \gamma_2 = \sqrt{i\omega r c_{\mathrm{tot}}}, \qquad Z_{c,2} = \sqrt{\frac{r}{i\omega c_{\mathrm{tot}}}}.$$

The channel is treated as open at both ends:

$$I_1(-L_1) = 0, \qquad I_2(L_2) = 0.$$

The boundary at $x = L_2$ is the contact-free end of the dual-gated region. The boundary at $x = -L_1$ is the graphite-contact end of the channel; its RF contact impedance is assumed to be much larger than the distributed channel impedance, so no appreciable RF conduction current flows into the contact.

The Region-1 impedance viewed from the inter-region boundary is

$$Z_{\mathrm{in},1} = Z_{c,1}\coth(\gamma_1 L_1).$$

Defining

$$\eta = \frac{Z_{\mathrm{in},1}}{Z_{c,2}}, \qquad E = e^{-2\gamma_2 L_2}, \qquad D(\eta) = (1+\eta) + (1-\eta)E,$$

the Region-2 profiles are

$$V_2(x) = \alpha V_A[1 - R_V(x)], \qquad I_2(x) = \frac{\alpha V_A}{Z_{c,2}} R_I(x),$$

with

$$R_V(x) = \frac{e^{-\gamma_2 x} + e^{-\gamma_2(2L_2 - x)}}{D(\eta)}, \qquad R_I(x) = \frac{e^{-\gamma_2 x} - e^{-\gamma_2(2L_2 - x)}}{D(\eta)}.$$

Continuity of voltage and current at $x = 0$ fixes the Region-1 profiles,

$$V_1(x) = V_2(0)\frac{\cosh[\gamma_1(x + L_1)]}{\cosh(\gamma_1 L_1)},$$

$$I_1(x) = \frac{V_2(0)}{Z_{c,1}}\frac{\sinh[\gamma_1(x + L_1)]}{\cosh(\gamma_1 L_1)}.$$

**Top-gate admittance**

The current drawn from the top-gate node by the distributed channel is

$$I_{\mathrm{TG}} = i\omega c_{\mathrm{TG}} \int_0^{L_2} [V_A - V_2(x)]\ dx.$$

The resulting distributed-line admittance viewed from the top gate is

$$Y_{\mathrm{line}}(\omega) = i\omega c_{\mathrm{TG}} \left[\beta L_2 + \frac{\alpha}{\gamma_2[\coth(\gamma_2 L_2) + \eta]}\right].$$

The total device admittance and impedance are

$$Y_{\mathrm{dev}}(\omega) = Y_{\mathrm{line}}(\omega) + i\omega C_{\mathrm{sh}}, \qquad Z_{\mathrm{dev}}(\omega) = \frac{1}{Y_{\mathrm{dev}}(\omega)}.$$

In the high-resistance limit,

$$Y_{\mathrm{dev}} \to i\omega(C_{\mathrm{sh}} + c_{\mathrm{TG}}\beta L_2) = i\omega\left(C_{\mathrm{sh}} + \frac{c_{\mathrm{TG}}c_{\mathrm{BG}}}{c_{\mathrm{TG}} + c_{\mathrm{BG}}} L_2\right) \qquad (R \to \infty).$$

This residual term is the local series capacitance from the top gate through an electrically isolated channel segment to the back gate.

In the low-resistance limit, the channel approaches an equipotential floating conductor. The effective top-gate-to-back-gate capacitance approaches

$$C_{\mathrm{line,on}} = \frac{(c_{\mathrm{TG}}L_2)[c_{\mathrm{BG}}(L_1 + L_2)]}{c_{\mathrm{TG}}L_2 + c_{\mathrm{BG}}(L_1 + L_2)}.$$

Because $C_{\mathrm{line,on}} > c_{\mathrm{TG}}\beta L_2$, the effective device capacitance is larger in the conducting limit and the resonance frequency is correspondingly lower.

**Matching network and resonance extraction**

The device is placed in series with an inductor $L_c$. Inductor loss is represented by

$$R_L(\omega) = \frac{\omega L_c}{Q_L}, \qquad Z_L = R_L + i\omega L_c,$$

where $Q_L$ = 95.5 is the inductor quality factor. The series branch, shunt-node admittance and full input impedance are

$$Z_{\text{series}} = Z_L + Z_{\text{dev}},$$

$$Y_{\text{node}} = \frac{1}{Z_{\text{series}}} + i\omega C_{\text{match}},$$

$$Z_{\text{in}} = \frac{1}{Y_{\text{node}}} + \frac{1}{i\omega C_{\text{block}}}.$$

The reflection coefficient at a $Z_0$ = 50 Ω input is

$$S_{11}(\omega) = \frac{Z_{\text{in}}(\omega) - Z_0}{Z_{\text{in}}(\omega) + Z_0}.$$

For each $R$, the simulated resonance frequency $f_0$ is obtained by fitting the complete complex $S_{11}(f)$ response of the distributed-RC device embedded in the full RF matching network to the complex-Lorentzian resonance form used for the measured traces in section 5. This definition keeps the simulated resonance extraction consistent with the experimental line-shape analysis.

**RF voltage, carrier modulation and dissipation**

The input-power parameter $P_{\text{in}}$ is the available incident power at the 50 Ω input port feeding the matching-network/sample assembly. The corresponding peak Thevenin source voltage is

$$V_{\text{src,pk}} = \sqrt{8 Z_0 P_{\text{in}}}.$$

The input current, shunt-node voltage and top-gate-node voltage are

$$I_{\text{in}} = \frac{V_{\text{src,pk}}}{Z_0 + Z_{\text{in}}}, \qquad V_{\text{node}} = \frac{I_{\text{in}}}{Y_{\text{node}}}, \qquad V_A = V_{\text{node}} \frac{Z_{\text{dev}}}{Z_{\text{series}}}.$$

The carrier-density modulation is calculated from the gate-capacitor charge balance. In Region 1,

$$\Delta n_1(x) = \frac{C_{\text{BG},A} V_1(x)}{e},$$

whereas in Region 2,

$$\Delta n_2(x) = \frac{C_{\mathrm{BG},A} V_2(x) + C_{\mathrm{TG},A}[V_2(x) - V_A]}{e}.$$

Panels S3d–f show peak-phase spatial profiles. For each quantity, the RF phase is chosen independently so that the largest absolute value of that quantity appears as a positive maximum in the plotted profile. The current $I(x)$ is the lateral current crossing the full channel width at position $x$; the plotted sheet current density, equivalent to the current per unit channel width, is $i_{\mathrm{AC}}(x) = I(x)/W$. Panel S3g shows the time-averaged Joule heating per unit channel length,

$$\langle p(x) \rangle_t = \frac{1}{2} |I(x)|^2 r.$$

**Numerical response and parameter range**

The calculation uses $L_c$ = 110 nH, $Q_L$ = 95.5, $C_{\mathrm{match}} = C_{\mathrm{block}}$ = 100 pF, $C_{\mathrm{sh}}$ = 6.285 pF, $Z_0$ = 50 Ω and $P_{\mathrm{in}}$ = −92 dBm. The modeled sheet resistance is swept over

$$10^1 \leq R \leq 10^8\ \Omega.$$

Over this range, the simulated resonance frequency increases from approximately 193.7 MHz in the conducting limit to approximately 194.5 MHz in the insulating limit. Evaluating each selected sheet resistance at its corresponding $f_0(R)$, the top-gate-node voltage remains below approximately 0.24 mV peak-to-peak. The maximum peak-phase lateral sheet current density is approximately 1.7 nA $\mu m^{-1}$, the maximum carrier-density modulation is approximately $0.65 \times 10^8$ cm$^{-2}$ and the maximum time-averaged Joule heating per unit channel length is approximately 0.016 pW $\mu m^{-1}$.

**Quantum-capacitance approximation**

Quantum-capacitance corrections are neglected. For a non-interacting parabolic two-dimensional band,

$$C_Q = \frac{e^2 g_d m_{\mathrm{eff}}}{2\pi\hbar^2}.$$

Here $m_{\mathrm{eff}}$ is the band effective mass. This gives $C_{\mathrm{Q,e}} \approx 2.17$ F m$^{-2}$ for the twelvefold-degenerate Q-valley electron band and $C_{\mathrm{Q,h}} \approx 0.84$ F m$^{-2}$ for the twofold-degenerate Γ-valley hole band. These values exceed the geometric capacitances,

$$C_{\mathrm{BG},A} \approx 2.66 \times 10^{-3}\ \mathrm{F\,m^{-2}}, \qquad C_{\mathrm{TG},A} \approx 2.04 \times 10^{-3}\ \mathrm{F\,m^{-2}},$$

by more than two orders of magnitude. The model therefore uses geometric hBN capacitances excluding quantum capacitance correction which is valid approximation.

The model above gives the resonance frequency $f_0(R)$ of the complete circuit for any sheet resistance $R$. The remaining step is to convert the measured $f_0$ into $R$. Because the full gate-induced swing of $f_0$ is only a few times larger than the slow temperature drift of the matching-network parameters, a direct inversion of $f_0(R)$ would be unreliable; the rest of this section therefore introduces a normalized frequency shift, $\chi$, that cancels these drifts, and describes the resulting resistance-extraction pipeline.

## The $\chi$ inversion method for channel resistance extraction

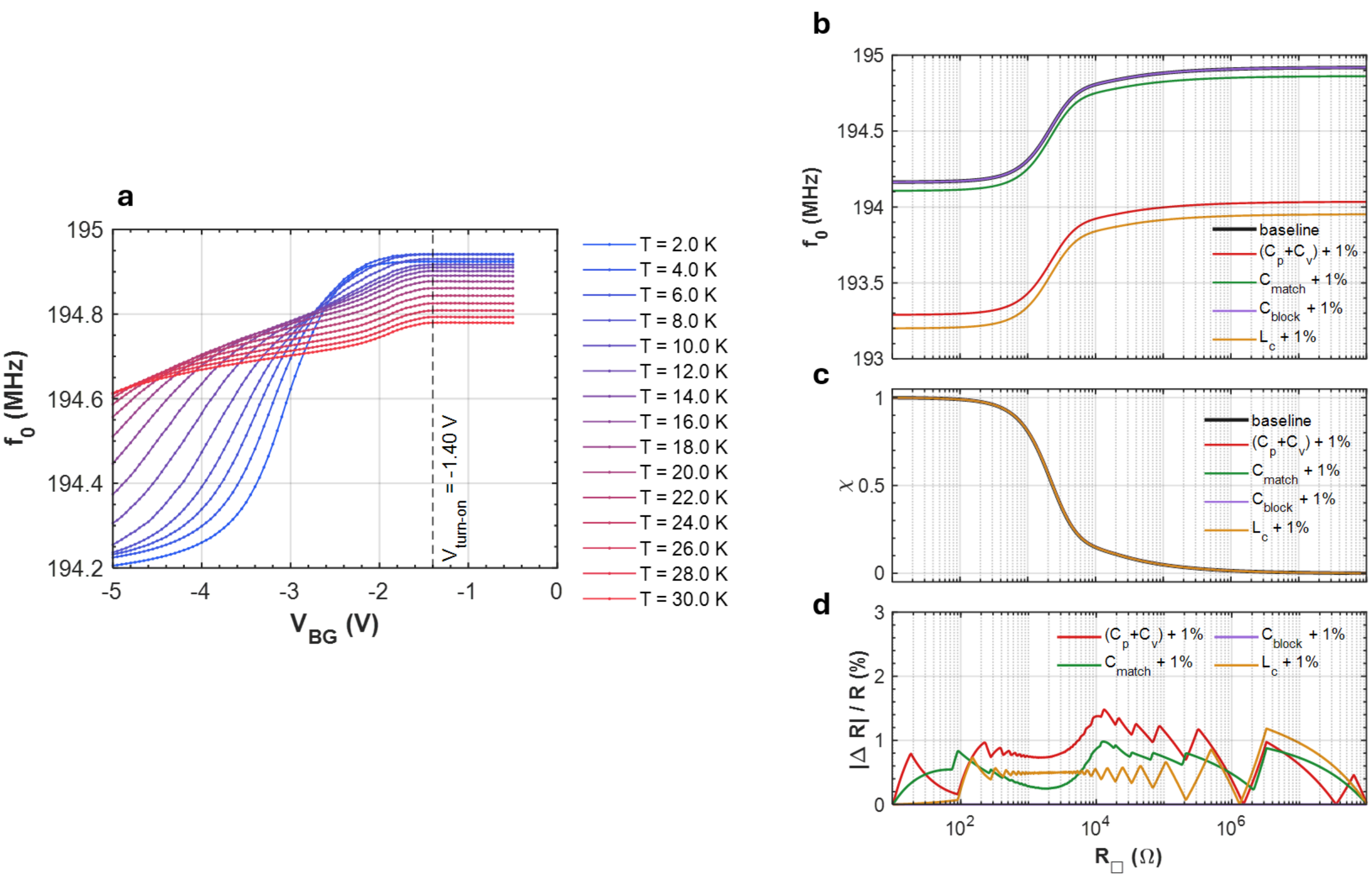

**Fig. S3. Robust extraction of channel sheet resistance from the resonance frequency via the $\chi$ inversion method.** **a**, Measured resonance frequency $f_0$ as a function of back-gate voltage $V_{BG}$ on the hole side at 15 temperatures spanning 2–30 K. The dashed vertical line marks the carrier-onset voltage $V_0 = -1.40$ V. The off-state $f_0$ plateau, measured to the right of $V_0$, drifts upward with increasing temperature by approximately 0.2 MHz over the 2–30 K window. This drift is small on the scale of the absolute resonance frequency but comparable to the gate-induced resonance-frequency swing. **b–d**, Numerical sensitivity test of the $\chi$ inversion against a representative +1% perturbation of each matching-network parameter, applied one at a time to $C_{sh} = C_p + C_v$, $C_{match}$, $C_{block}$ and $L_c$. **b**, Simulated $f_0(R)$ for the baseline matching network and for each perturbed parameter, using the same distributed-RC device admittance of this section embedded in the full matching network of section 2. For each simulated trace, $f_0$ is obtained by fitting the complete complex $S_{11}(f)$ response of the full circuit to the complex-Lorentzian resonance form used for the measured traces. **c**, $\chi(R)$ computed from the same $f_0(R)$ curves using Eq. (S3.1) and each curve's own ($f_{off}$, $f_{on}$) anchors. The curves overlay to plotting precision. **d**, Relative resistance-extraction error $|\Delta R|/R$ obtained by inverting the baseline $\chi$ curve through a $\chi_{model}$ curve calculated with one +1%-perturbed parameter. The error remains below approximately 1.5% over the full modeled resistance range.

The off-state resonance frequency $f_{off}$ drifts by approximately 0.1% of the absolute resonance frequency over the 2–30 K range (Fig. S3a). Although this absolute drift is small, the full gate-induced frequency swing between the insulating and conducting channel limits is only about 0.4% of $f_0$. A direct lookup of $R$ from the absolute value of $f_0$ would therefore be sensitive to small temperature-dependent changes in the parasitic capacitances and lumped matching-network components. To avoid this uncertainty, we use a dimensionless normalized frequency shift, $\chi$, defined from the measured resonance frequency and its insulating and conducting anchors. This normalization is the quantity inverted to obtain $R$ in the main text.

**Definition of $\chi$**

For each gate-voltage sweep, we define

$$\chi(V_{\mathrm{BG}}) = \frac{1/f_0(V_{\mathrm{BG}})^2 - 1/f_{\mathrm{off}}^2}{1/f_{\mathrm{on}}^2 - 1/f_{\mathrm{off}}^2} \quad \text{(S3.1)}$$

Here $f_{\mathrm{off}}$ is the resonance frequency in the insulating limit, where $R \to \infty$, and $f_{\mathrm{on}}$ is the resonance frequency in the conducting limit, where $R \to 0$. By construction, $\chi = 0$ in the insulating limit and $\chi = 1$ in the conducting limit.

The use of $1/f_0^2$ follows from the high-$Q$, lossless tank limit, in which

$$(2\pi f_0)^2 \propto \frac{1}{L_c[C_{\mathrm{sh}} + C_{\mathrm{dev}}(R)]}.$$

Here $C_{\mathrm{sh}} = C_p + C_v$ is the resistance-independent device-side shunt capacitance and $C_{\mathrm{dev}}(R)$ is the resistance-dependent effective capacitance of the distributed channel. In this limit, the normalization in Eq. (S3.1) cancels $L_c$ and $C_{\mathrm{sh}}$.

**Numerical demonstration of $\chi$ cancellation**

To quantify the residual sensitivity to circuit-parameter drift, we repeat the full distributed-RC plus matching-network simulation for the baseline circuit and for four perturbed circuits. In the perturbed simulations, one parameter at a time is increased by 1%: $C_{\mathrm{sh}}$, $C_{\mathrm{match}}$, $C_{\mathrm{block}}$ or $L_c$. All other parameters are held fixed.

For each circuit simulation, $R$ is swept over the same logarithmic grid and the full complex reflection response $S_{11}(f)$ is calculated. The simulated resonance frequency $f_0(R)$ is extracted by fitting each complex $S_{11}(f)$ trace to the complex-Lorentzian resonance form used for the measured traces in section 5. The resulting $f_0(R)$ curves shift visibly under 1% perturbations of the matching-network parameters (Fig. S3b).

For each simulated curve we then form $\chi(R)$ using Eq. (S3.1), taking $f_{\mathrm{off}}$ and $f_{\mathrm{on}}$ from that same curve. This normalization removes the apparent circuit-parameter drift: the resulting $\chi(R)$ curves overlay to plotting precision (Fig. S3c). To quantify the remaining error, the baseline $\chi(R)$ curve is inverted through each perturbed $\chi_{\mathrm{model}}(R)$ curve. The resulting relative resistance error,

$$\frac{|\Delta R|}{R},$$

remains below approximately 1.5% across the modeled range (Fig. S3d).

### Extraction of $f_{\mathrm{off}}(T)$ and $f_{\mathrm{on}}(T)$

The two anchors are obtained directly from each $V_{\mathrm{BG}}$ sweep where possible. The off-state anchor $f_{\mathrm{off}}(T)$ is read from the depletion plateau, where $R$ has exceeded the RF-sensitive range and $f_0$ has reached its high-resistance limit. The on-state anchor $f_{\mathrm{on}}(T)$ is taken from the opposite end of the sweep, where the channel has entered the low-resistance plateau.

At the base temperature $T_0 = 2$ K, both anchors are accessible within the measured gate-voltage range. At higher temperatures, the maximum applied $|V_{\mathrm{BG}}|$ does not always drive the channel fully into the low-resistance plateau. For $T > T_0$, we therefore use the base-temperature on–off separation as a geometric calibration of the total channel-capacitance swing:

$$\Delta f_{\mathrm{geom}} = f_{\mathrm{off}}(T_0) - f_{\mathrm{on}}(T_0),$$

and define

$$f_{\mathrm{on}}(T) = f_{\mathrm{off}}(T) - \Delta f_{\mathrm{geom}} \quad \text{(S3.2)}$$

This procedure is valid because gate-capacitance swing of the device is set primarily by the hBN dielectric constant and channel geometry which is temperature independent over the measurement range ($T = 2 \sim 30$ K).

### Resistance-extraction pipeline

For each temperature and magnetic field, the analysis proceeds as follows.

1. The measured complex $S_{11}(f)$ traces are fitted with the global line-shape procedure described in section 5, giving $f_0(V_{\mathrm{BG}})$.

2. The off-state anchor $f_{\mathrm{off}}(T)$ is read from the deep-depletion plateau.

3. The on-state anchor $f_{\mathrm{on}}(T)$ is read directly when the conducting plateau is reached, or assigned using Eq. (S3.2) when the conducting plateau is not fully exposed.

4. The measured $\chi_{\mathrm{meas}}(V_{\mathrm{BG}})$ is calculated from Eq. (S3.1).

5. The channel sheet resistance is obtained by interpolating $\chi_{\mathrm{meas}}$ through the model curve $\chi_{\mathrm{model}}(R)$ calculated from the full complex-$S_{11}$ distributed-RC simulation.

**Parameter values used in Fig. S3**

Panel a shows the measured hole-side $V_{\mathrm{BG}}$ sweep for $T = 2, 4, 6, \ldots, 30\ K$. Panels b–d use the baseline matching-network values

$$C_{\mathrm{sh}} = C_p + C_v = 6.285\ \mathrm{pF}, \quad L_c = 110\ \mathrm{nH}, \quad Q_L = 95.5,$$

$$C_{\mathrm{match}} = C_{\mathrm{block}} = 100\ \mathrm{pF}, \quad Z_0 = 50\ \Omega,$$

and the device geometry given above:

$$W = 11\ \mu\mathrm{m}, \quad L_1 = 16\ \mu\mathrm{m}, \quad L_2 = 14\ \mu\mathrm{m},$$

$$t_{\mathrm{BG}} = 10\ \mathrm{nm}, \quad t_{\mathrm{TG}} = 13\ \mathrm{nm}, \quad \epsilon_{\mathrm{hBN}} = 3.$$

The model sheet-resistance grid is logarithmic, spanning $10^1 \leq R \leq 10^8\ \Omega$.

## Section 4: Gate-tuned phase diagram

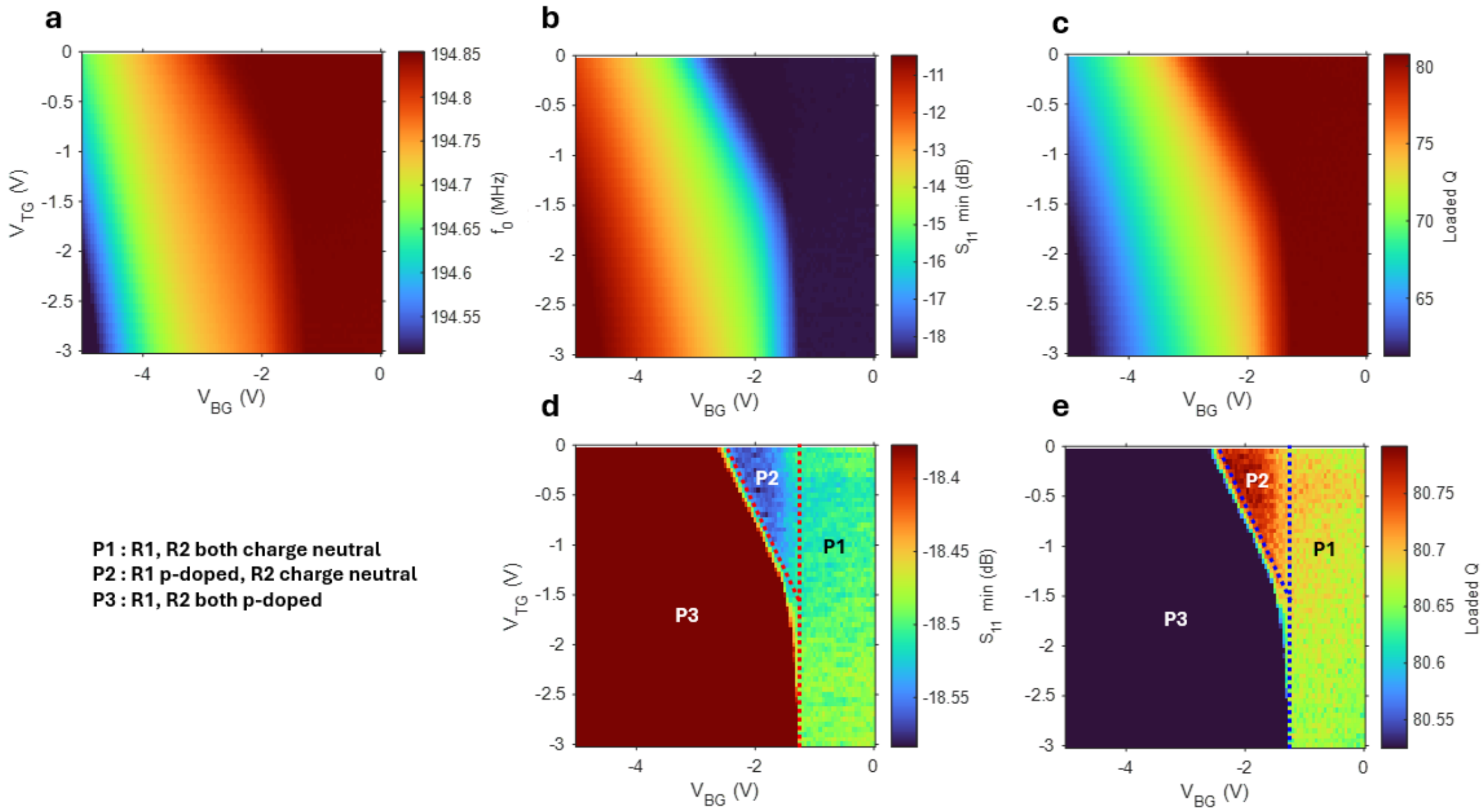

**Fig. S4. Gate-tuned phase diagram of the dual-region 2L-$MoSe_2$ channel at $T$ = 20 K (hole side). a–c**, Two-dimensional maps across the ($V_{BG}$, $V_{TG}$) plane of (**a**) resonance frequency $f_0$, (**b**) minimum reflection magnitude min|$S_{11}$|, and (**c**) quality factor $Q$, extracted at each grid point by an independent local line-shape fit (Stage 1 of the section 5 pipeline). **d**,**e**, Same data as **b** and **c**, respectively, with the color scale restricted around the off-state values to expose three regions P1, P2, and P3 separated by dashed boundaries (guides to the eye drawn along the loci of most abrupt change). P1: Region 1 and Region 2 both at charge neutrality. P2: Region 1 p-doped, Region 2 at charge neutrality. P3: Region 1 and Region 2 both p-doped.

The top-gate metal is deposited on Region 2 only. Its work-function difference with 2L-$MoSe_2$ imposes a built-in surface potential on Region 2 even at $V_{TG}$ = 0, leaving Region 2 with a smaller charge density than Region 1 at any given $V_{BG}$. A finite $V_{TG}$ compensates this offset; we map the line-shape response in the ($V_{BG}$, $V_{TG}$) plane and identify the $V_{TG}$ at which Regions 1 and 2 reach charge neutrality at the same $V_{BG}$. The data shown in Fig. S4 are on the hole side; the same strategy applies on the electron side.

**Measurement.** At each grid point in ($V_{BG}$, $V_{TG}$) we record an $S_{11}(f)$ trace at $T$ = 20 K and extract $f_0$, min|$S_{11}$|, and $Q$ by an independent local fit (Fig. S4a–c). Region 1 and Region 2 phase boundaries are sharpest in panels Fig. S4d,e where the color scale is restricted around the off-state values.

**Phase diagram identification.** Three regions are visible in the ($V_{BG}$, $V_{TG}$) plane. In P1, near the weakly negative-$V_{BG}$ side of the map, both regions are depleted and their resistance exceeds the RF-sensitive range. In P3, both regions are p-doped and the channel has finite conductance, allowing the RF response to penetrate into Region 1 from the top-gated region. In the intermediate strip P2, Region 1 is p-doped while Region 2 remains near charge neutrality. The three observables vary continuously across the plane; the dashed boundaries in Fig. S4d,e are guides to the eye drawn along the loci of most abrupt change.

**Absence of a P4 phase.** The complementary configuration — Region 1 charge-neutral and Region 2 p-doped — was not observed in this sweep geometry because depletion of Region 1 interrupts the conductive path from the graphite electrode to Region 2.

**Boundary geometry.** The P1/P2 boundary is approximately vertical because Region 1 is gated by $V_{BG}$ alone. The P2/P3 boundary is diagonal because Region 2 is dual-gated; its slope $dV_{BG}/dV_{TG}$ is set by the ratio of the top- and back-gate capacitances per unit area.

**Carrier-density matching.** The back-gate dielectric is uniform along the channel, so Regions 1 and 2 share the same back-gate capacitance per unit area. Sweeping $V_{BG}$ produces the same $dn/dV_{BG} = C_{BG}/e$ in both regions at any fixed $V_{TG}$; $V_{TG}$ sets only the relative threshold offset. Locking $V_{TG}$ to the value at which the P1/P2 and P2/P3 boundaries cross places both regions at charge neutrality at the same $V_{BG}$, and the channel remains density matched as $V_{BG}$ is swept. The resulting $V_{TG}$ — 0.45 V on the electron side and −1.5 V on the hole side — is used for all low-temperature $V_{BG}$ sweeps in the main text; the contact gate is held at $V_{CG}$ = +6 V and −6 V, respectively, to maintain rapid electron and hole injection during gate sweeps.

**Parameter values used in Fig. S4.** $V_{BG}$: 0 to −5 V; $V_{TG}$: 0 to −3 V; $T$ = 20 K; $P_{in}$ = −92 dBm; contact gate held at $V_{CG}$ = −6 V to maintain rapid hole injection from the graphite contact during the gate sweep.

## Section 5: Global line-shape fitting

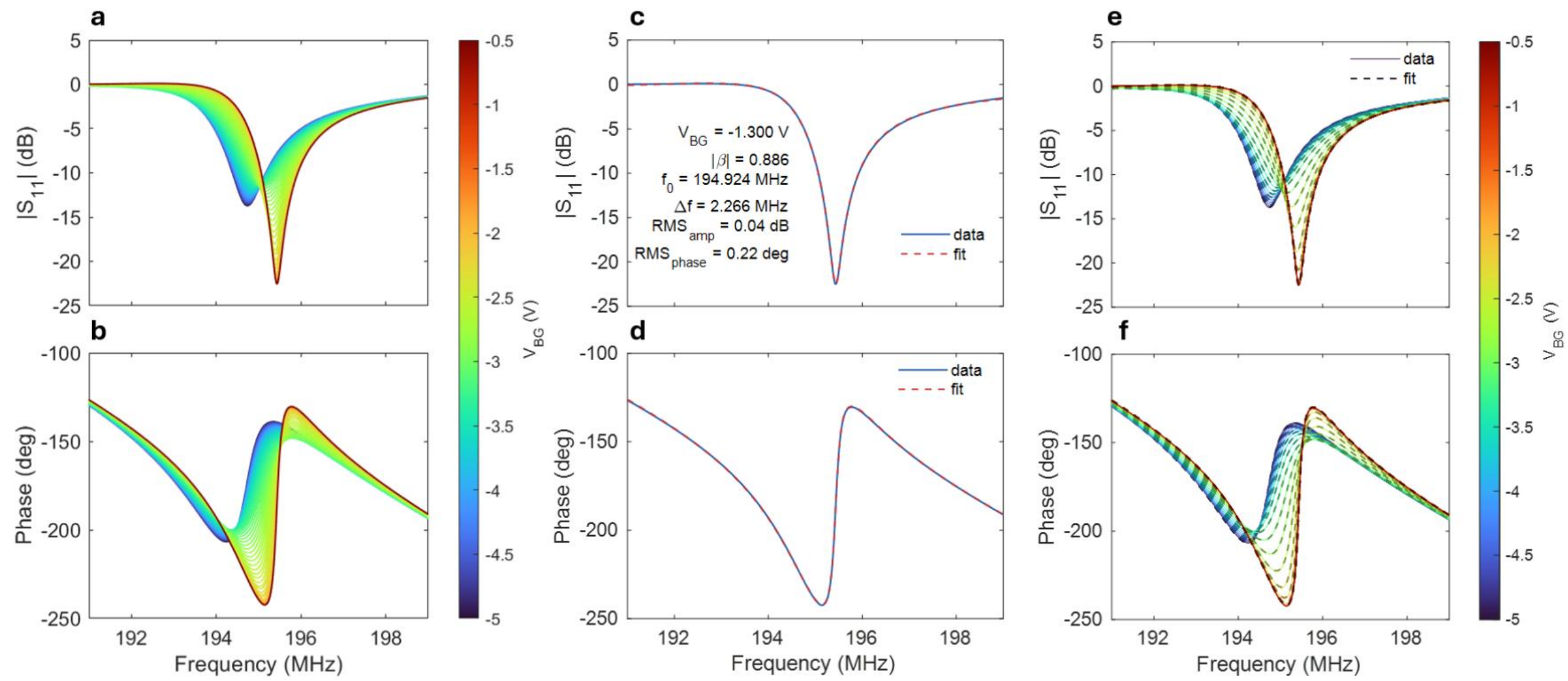

**Fig. S5. Global line-shape fit across the back-gate sweep, hole side at *T* = 2 K. a,b**, Amplitude $|S_{11}|(f)$ (**a**) and phase($f$) (**b**) measured at every $V_{BG}$ in the sweep; $V_{BG}$ is mapped to color by the right-hand color bar, which applies to every colored trace in this figure. **c,d**, Measurement (solid navy) and global-fit model (dashed crimson) at one representative $V_{BG}$, for $|S_{11}|(f)$ (**c**) and phase($f$) (**d**). The annotation in **c** reports $V_{BG}$, the local parameters $|\beta|$, $f_0$ and $\Delta f$, and the per-trace residuals $RMS_{amp}$ (dB) and $RMS_{phase}$ (deg). **e,f**, Same data–model comparison as in **c,d**, overlaid for every fourth $V_{BG}$. The shared-baseline model reproduces the amplitude and phase line shapes across the sweep.

**Line-shape model.** We describe each measured $S_{11}(f)$ trace as a complex Lorentzian on a real off-resonance background, with linear amplitude and phase baselines that absorb residual instrument drift:

$$S_{11}(f) = \alpha_{re} + (\beta_{re} + i\beta_{im}) / (1 + ix), \quad x = 2(f - f_0) / \Delta f,$$

$$|S_{11}|_{dB}(f) = 20 \log_{10} |S_{11}(f)| + b_0 + b_1 (f - f_c),$$

$$\text{phase}(f) = \arg S_{11}(f) \cdot (180/\pi) + \varphi_0 + \varphi_1 (f - f_c),$$

where $f_c$ is the midpoint of the measured frequency window. The complex residue $\beta = \beta_{re} + i\beta_{im}$ sets the depth and asymmetry of the dip; $\alpha_{re}$ is the off-resonance reflection floor; $b_0$, $b_1$ remove a constant offset and a linear slope in dB across the band, and $\varphi_0$, $\varphi_1$ play the same role for the phase. The resonance frequency $f_0$ and full width at half maximum $\Delta f$ define the loaded quality factor $Q = f_0 / \Delta f$.

**Shared and local parameters.** In the device off-state the asymmetry phase $\theta_\beta \equiv \arg(\beta)$ is dominated by a fixed cable artifact rather than by any $V_{BG}$-dependent device response. We therefore lock $\theta_\beta$ across the sweep and allow only its magnitude $|\beta|$ to vary with the gate. The parameter vector then separates into a **shared** group $\alpha_{re}$, $b_0$, $b_1$, $\varphi_0$, $\varphi_1$ and $\theta_\beta$ (six parameters, fit once per sweep) and a **local** group $|\beta|$, $f_0$ and $\Delta f$ (three parameters per gate point). For a sweep of $N_V$ gate points the total number of free parameters is therefore $6 + 3N_V$.

**Two-stage optimization.** Stage 1 fits each $V_{BG}$ independently with all parameters free, providing robust initial guesses for $f_0$, $\Delta f$ and the baselines. Stage 2 then minimizes the joint amplitude-plus-phase residual

$$\chi^2 = \Sigma V_{BG,f} \left[ (|S_{11}|_{dB}{}^{data} - |S_{11}|_{dB}{}^{model})^2 + w^2 (\text{phase}^{data} - \text{phase}^{model})^2{}_{wrap} \right],$$

over all $6 + 3N_V$ parameters simultaneously. The wrap operator on the phase residual handles the 360° branch cut, and $w = 8.686 \cdot \pi/180$ converts degrees of phase residual into an equivalent dB amplitude, placing the two channels on a common footing. The minimization uses the Levenberg–Marquardt algorithm (MATLAB lsqnonlin) with parameter bounds set wide enough to avoid edge solutions. Per-trace fit quality is reported by $\text{RMS}_{amp} = \sqrt{\langle(|S_{11}|_{dB}{}^{data} - |S_{11}|_{dB}{}^{model})^2\rangle}$ and $\text{RMS}_{phase} = \sqrt{\langle(\text{phase}^{data} - \text{phase}^{model})^2{}_{wrap}\rangle}$, evaluated over the measurement frequency window at each $V_{BG}$.

**Section 6: Density and temperature dependence of the carrier mobility**

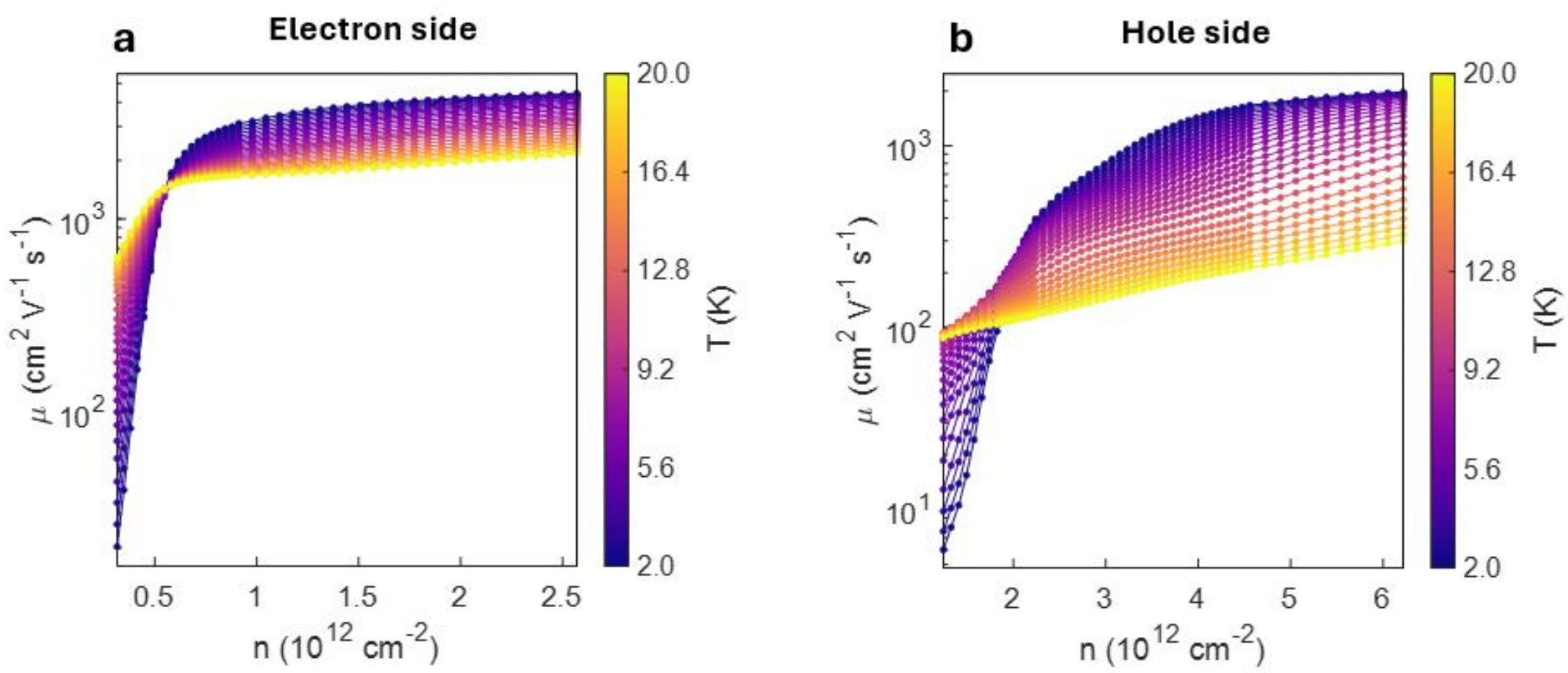

**Fig. S6. Density and temperature dependence of the mobility. a,b**, Mobility $\mu = 1/(enR)$, extracted from the contact-free RF sheet resistance, as a function of carrier density $n$ for electrons (a) and holes (b) at zero magnetic field. Colors denote temperature from $T = 2$ to 20 K. At $T = 2.0$ K, the maximum metallic-regime mobilities are $\mu_e^{max} \approx 4.44 \times 10^3$ cm$^2$ V$^{-1}$ s$^{-1}$ at $n \approx 2.57 \times 10^{12}$ cm$^{-2}$ and $\mu_h^{max} \approx 1.97 \times 10^3$ cm$^2$ V$^{-1}$ s$^{-1}$ at $n \approx 6.22 \times 10^{12}$ cm$^{-2}$. At the critical densities used in the Ioffe–Regel–Mott comparison, the corresponding mobilities are $\mu_e(n_c) \approx 940$ cm$^2$ V$^{-1}$ s$^{-1}$ at $n_c \approx 0.53 \times 10^{12}$ cm$^{-2}$ and $\mu_h(n_c) \approx 237$ cm$^2$ V$^{-1}$ s$^{-1}$ at $n_c \approx 1.98 \times 10^{12}$ cm$^{-2}$. These values are used to compare the measured ambipolar critical densities with a common Ioffe–Regel–Mott crossover criterion.

Figure S6 shows the carrier mobility $\mu = 1/(enR)$, obtained from the contact-free RF sheet resistance $R$ at zero magnetic field, as a function of carrier density $n$ for electrons and holes at temperatures from 2 to 20 K. The mobilities at the measured critical densities and the maximum metallic-regime mobilities quoted in the main text and section 1 are read from these curves at $T = 2.0$ K; they enter the Ioffe–Regel–Mott comparison of the two critical densities.

## Section 7: Device reproducibility

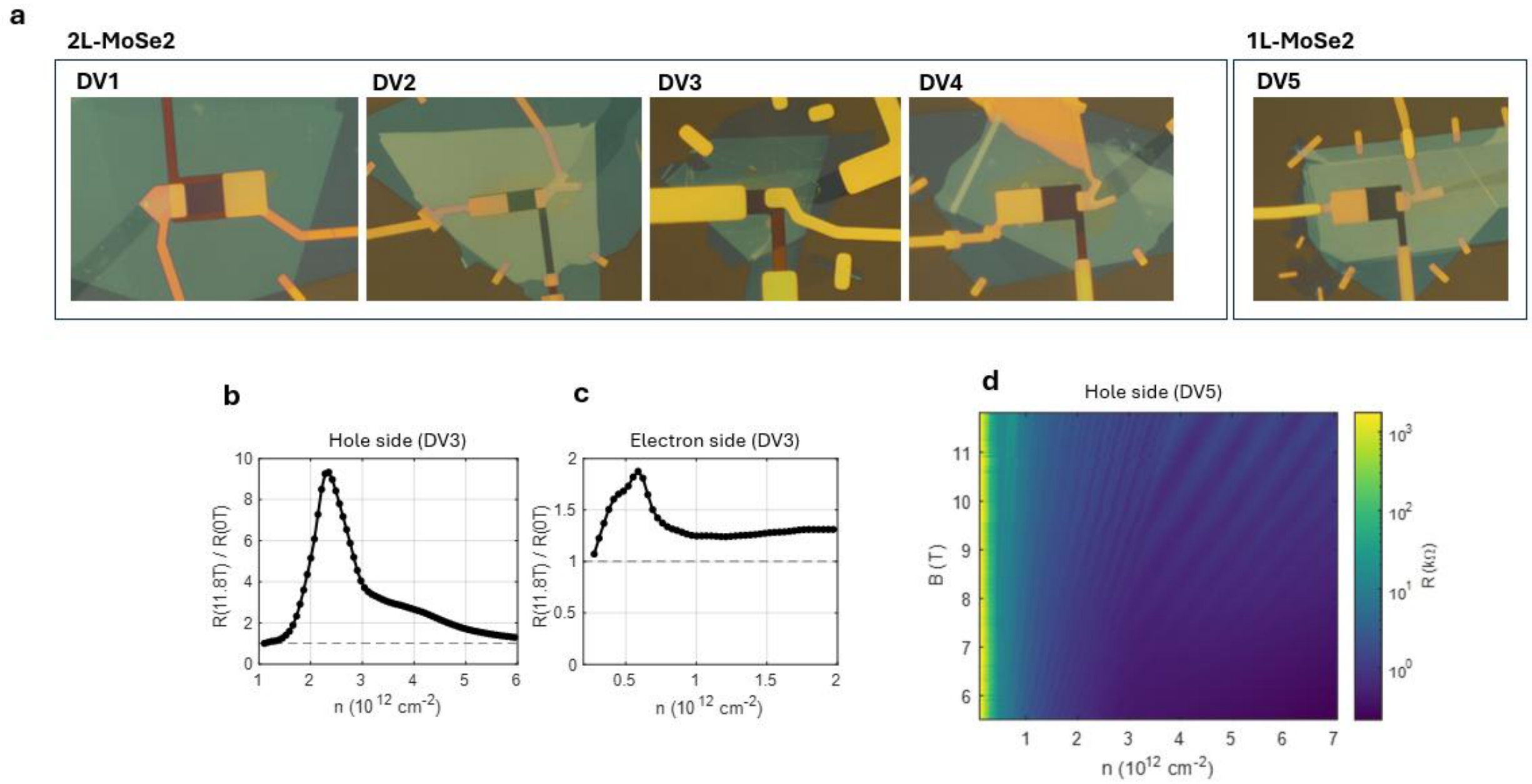

**Fig. S7. Device reproducibility.** **a**, Optical micrographs of the five measured devices: bilayer $MoSe_2$ devices DV1–DV4 and monolayer $MoSe_2$ device DV5. DV1 is the primary device of the main text. **b,c**, Magnetoresistance ratio $R(11.8\ \mathrm{T})/R(0)$ versus carrier density $n$ for device DV3 at $T = 2.5$ K in an in-plane field, on the hole side (**b**) and the electron side (**c**). Horizontal dashed lines mark $R(11.8\ \mathrm{T})/R(0) = 1$. **d**, Hole-side channel resistance $R$ of the monolayer device DV5 in the $(n, B_\perp)$ plane at $T = 2.5$ K, on a logarithmic color scale, showing Shubnikov–de Haas oscillations.

We fabricated and measured five $MoSe_2$ devices with the same two-region gate geometry and contact-free RF reflectometry scheme, four bilayer (DV1–DV4) and one monolayer (DV5) (Fig. S7a). DV1 is the primary device analyzed in the main text. The resonator couples capacitively to both channel regions, with a resistance-dependent penetration length, and the resonance-frequency response $f_0(V_{BG})$ therefore provides a sensitive diagnostic of macroscopic channel homogeneity. When the channel accumulates carriers uniformly, $f_0$ evolves smoothly with $V_{BG}$ as the whole channel turns on together. In devices with substantial macroscopic structural disorder, such as accumulated strain or trapped bubbles visible in the optical micrographs, $f_0(V_{BG})$ instead develops a staircase-like profile. Separate portions of the channel become conducting only above different threshold voltages, and the conducting area grows in discrete steps rather than continuously, indicating macroscopic transport inhomogeneity.

The two carrier polarities are not affected equally. In one device the electron side showed a smooth, near-ideal frequency response while the staircase behavior appeared only on the hole side. In the most inhomogeneous devices, with a high density of bubbles, both polarities show staircase responses. The two polarities are measured in the same channel and experience the same structural disorder. The asymmetry therefore indicates that the effect of a given strain profile on carrier mobility depends on the underlying band structure, with the Γ-point valence band and the Q-point conduction band of bilayer $MoSe_2$ coupling differently to the same local strain.

Devices DV1 and DV3 show smooth frequency responses on both the electron and hole sides, indicating that macroscopic disorder is small in these devices, and both were characterized in magnetotransport. In DV3 the magnetoresistance peak at the critical density is reproduced for both carrier types. Figure S7b,c show the in-plane-field ratio $R(11.8\ \mathrm{T})/R(0)$ at $T = 2.5$ K. The hole-side peak reaches ≈ 9.3 at $n \approx 2.3 \times 10^{12}$ cm$^{-2}$ and the electron-side peak reaches ≈ 1.8 at $n \approx 0.59 \times 10^{12}$ cm$^{-2}$, close to the primary-device values of $n_c \approx 1.98 \times 10^{12}$ cm$^{-2}$ for holes and ≈ $0.53 \times 10^{12}$ cm$^{-2}$ for electrons. The ratio of the two peak densities, ≈ 3.9, also reproduces the factor-of-four ambipolar ordering. The peak amplitude varies between devices, with a stronger hole-side peak in DV3 than in DV1, but the existence, sharpness and position of the peak are reproduced.

The technique also extends beyond the bilayer. The monolayer device DV5 exhibits higher mobility than the bilayer devices, sufficient to support quantum oscillations at the accessible magnetic fields. Figure S7d shows the hole-side resistance of DV5 in the $(n, B_\perp)$ plane at $T = 2.5$ K, in which Shubnikov–de Haas oscillations are clearly resolved. Low-temperature ohmic contact to the $MoSe_2$ valence band is particularly difficult, and the capacitive coupling circumvents this contact problem entirely. Hole-side quantum oscillations in a monolayer therefore demonstrate the applicability of the measurement across layer number and carrier polarity.

**Section 8: Field and temperature evolution of the resistance across the metal–insulator crossover**

This section presents the temperature and magnetic-field evolution of the sheet resistance across the metal–insulator crossover. Figure S8 shows the sheet resistance $R(T)$ at $B_\parallel = 0$ and 11.8 T for both carrier types. Figure S9 shows the hole-side $R(B_\parallel)$ traces at temperatures between 2.5 and 6.0 K, together with the crossover field $B^*(n)$, defined as the field at which the low-temperature derivative $dR/dT$, evaluated over 2.2–4 K, changes sign. Figure S10 shows the field evolution of the magnetoresistance ratio versus density for both carriers: the critical-density peak grows with field while remaining fixed near $n_c$, whereas the broader high-density maximum on the hole side shifts to higher density with field.

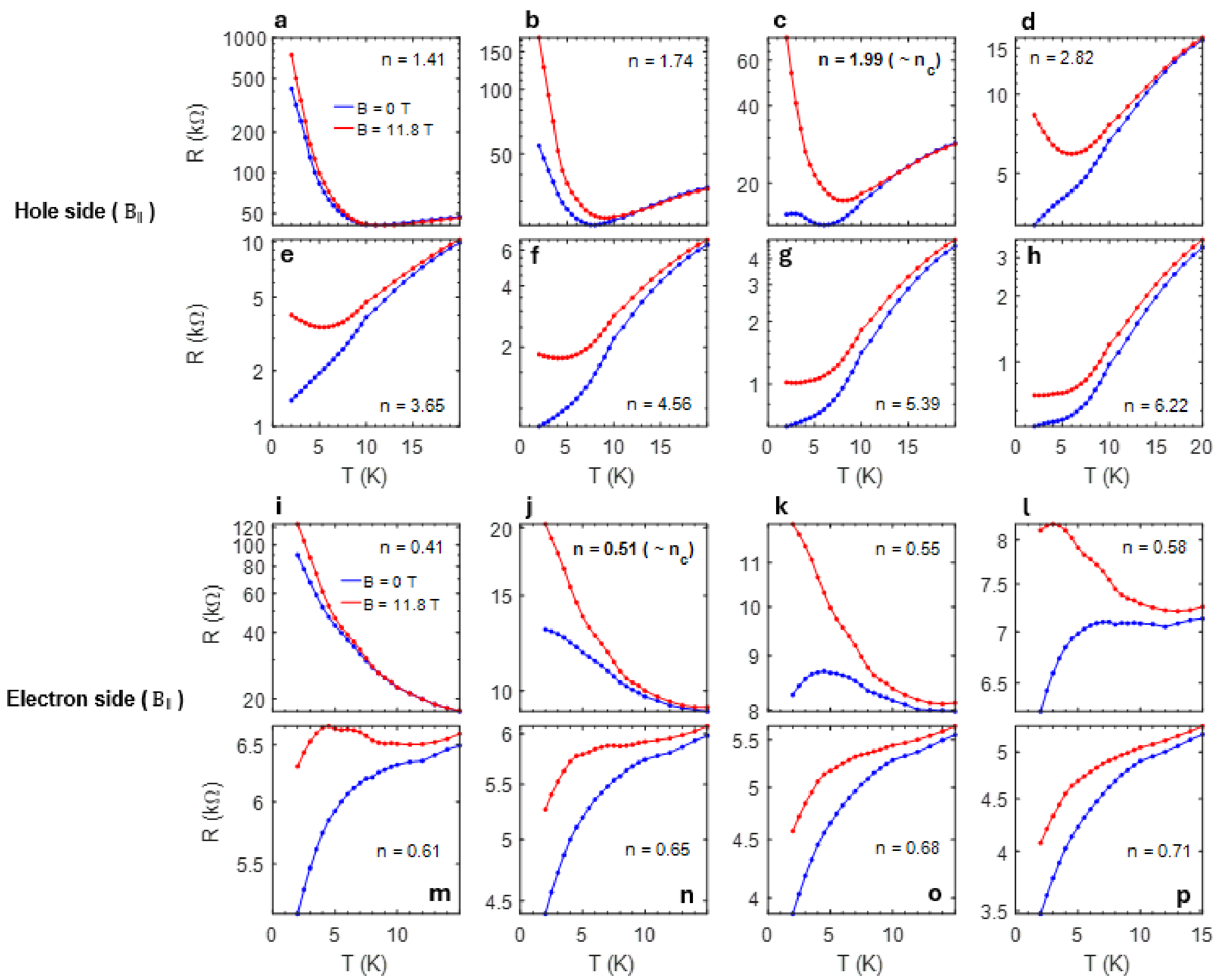

**Fig. S8. Temperature dependence of the resistance at zero and high in-plane field across the crossover.** Sheet resistance $R(T)$ at $B_\parallel = 0$ (blue) and 11.8 T (red). **a–h,** Hole side at $n$ = 1.41, 1.74, 1.99 ($\approx n_c$), 2.82, 3.65, 4.56, 5.39, and 6.22 × $10^{12}$ cm$^{-2}$, respectively. **i–p,** Electron side at $n$ = 0.41, 0.51 ($\approx n_c$), 0.55, 0.58, 0.61, 0.65, 0.68, and 0.71 × $10^{12}$ cm$^{-2}$, respectively. On the hole side below $n_c$, $R$ rises strongly on cooling. Above $n_c$, the high-field curves at intermediate densities remain metallic-like at higher temperatures but turn upward at low temperatures; this low-temperature upturn weakens toward higher density.

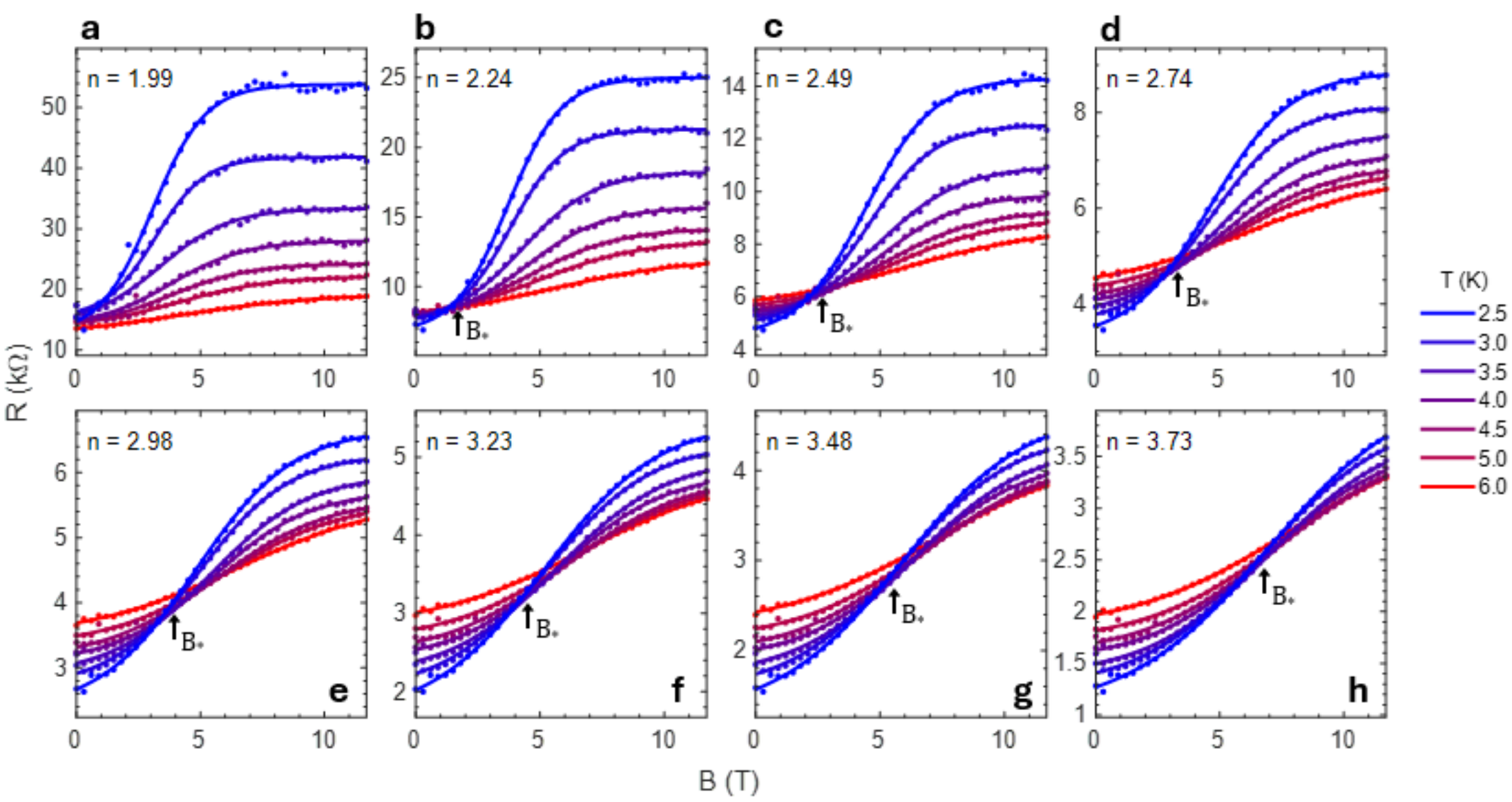

**Fig. S9. Density evolution of the temperature-dependent in-plane magnetoresistance on the hole side. a–h,** Sheet resistance $R(B)$ versus in-plane magnetic field $B_\parallel$ at $T = 2.5$, 3.0, 3.5, 4.0, 4.5, 5.0 and 6.0 K for hole densities $n = 1.99$, 2.24, 2.49, 2.74, 2.98, 3.23, 3.48 and $3.73 \times 10^{12}$ cm$^{-2}$, respectively. Dots are measured data and solid lines are logistic fits. Arrows in **b–h** mark the crossover field $B^*(n)$, at which the low-temperature slope changes from metallic-like ($dR/dT > 0$) at lower field to insulating-like ($dR/dT < 0$) at higher field. Panel **a** lies near the zero-field critical density, where the crossover occurs at $B^* \approx 0$; $B^*(n)$ increases with density.

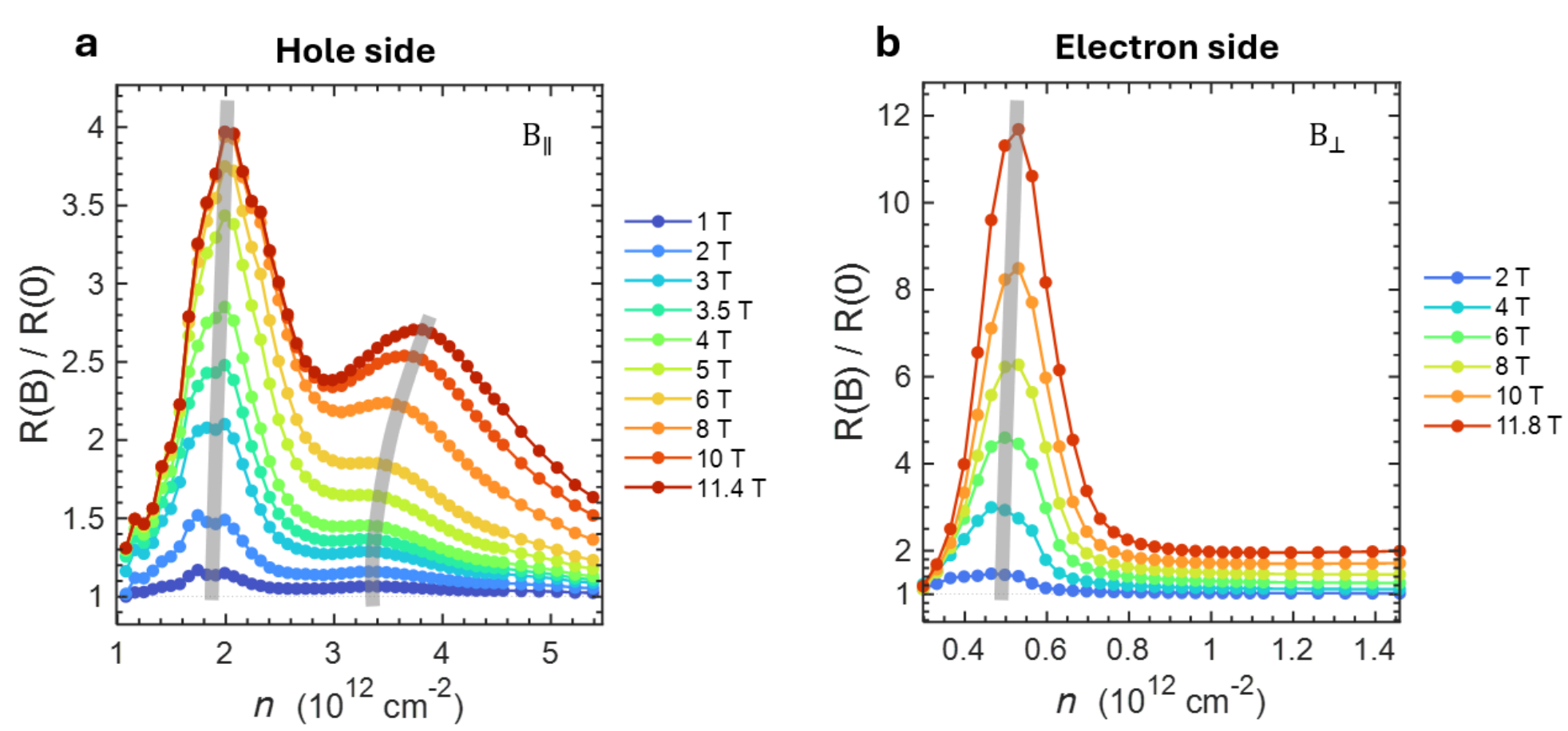

**Fig. S10. Field evolution of the magnetoresistance versus density. a**, Hole side (in-plane, $B_\parallel$): magnetoresistance ratio $R(B)/R(0)$ at $T = 2.2$ K as a function of carrier density $n$ for fields from 1 to 11.4 T. Two maxima strengthen with field but behave differently: the zero-field critical-density peak near $n \approx 2 \times 10^{12}$ cm$^{-2}$ remains approximately fixed in density, whereas the broader high-density peak near $n \approx 3.5$–$4.2 \times 10^{12}$ cm$^{-2}$ shifts to higher density with field. **b**, Electron side (out-of-plane, $B_\perp$): $R(B)/R(0)$ at $T = 2$ K as a function of $n$ for fields from 2 to 11.8 T; the zero-field critical-density peak near $n \approx 0.53 \times 10^{12}$ cm$^{-2}$ grows strongly while remaining approximately fixed in density, reaching $R(B)/R(0) \approx 12$ at 11.8 T. Gray lines are guides to the eye for the peak position as a function of field.

## Section 9: Saturated magnetoresistance ratio in dilute silicon MOSFETs

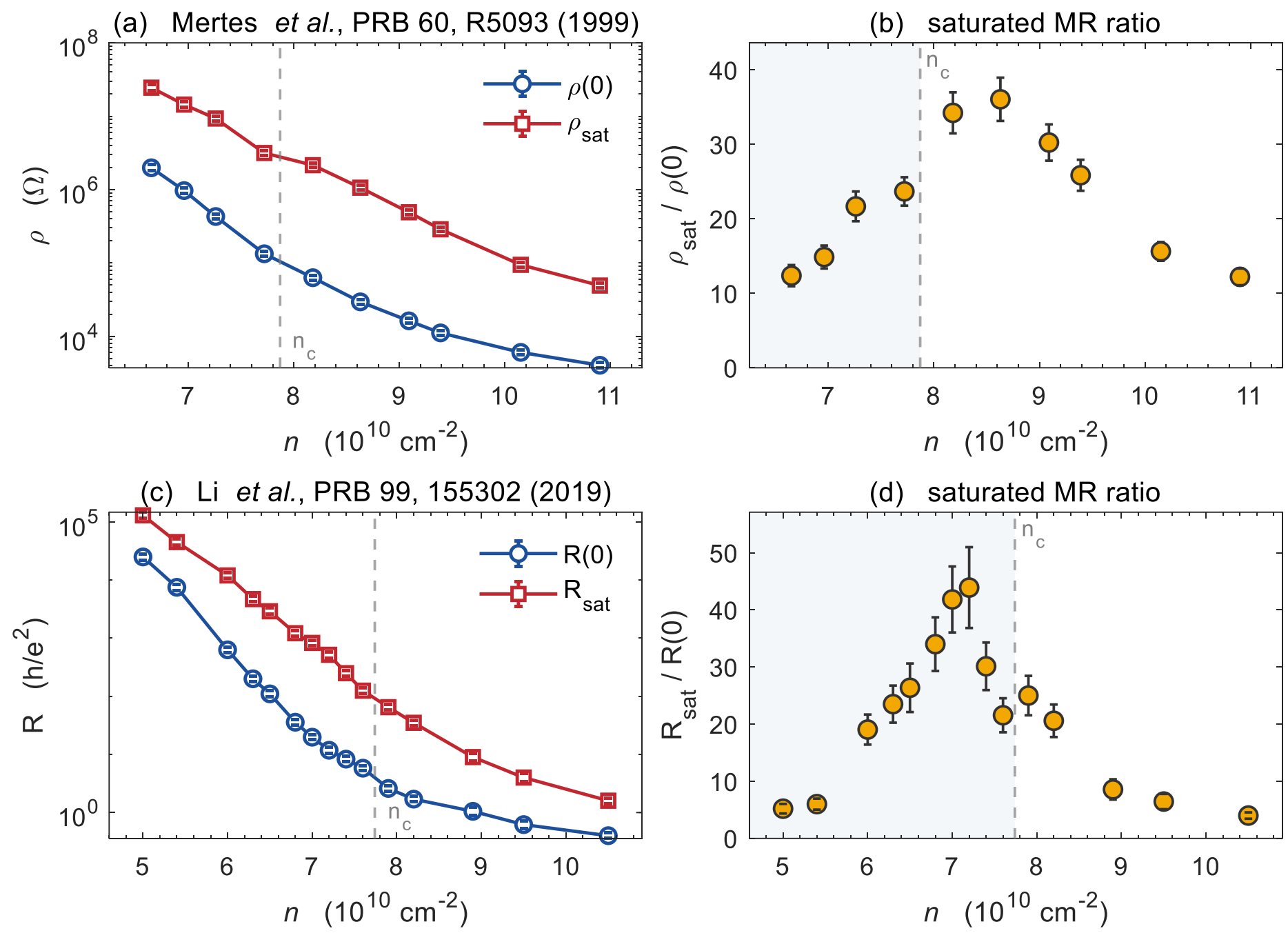

**Fig. S11. Saturated parallel-field magnetoresistance ratio in dilute silicon MOSFETs (digitized). a,b**, Mertes *et al.* (PRB 60, R5093 (1999)). **a**, Zero-field resistivity $\rho(0)$ and saturated high-field resistivity $\rho_{sat}$ versus carrier density $n$. **b**, Saturated magnetoresistance ratio $\rho_{sat}/\rho(0)$ versus $n$. **c,d**, Li *et al.* (PRB 99, 155302 (2019)), the same quantities with $R$ in units of $h/e^2$. Vertical dashed lines mark the critical densities $n_c$ reported in each work, and shading marks the insulating side. All points are digitized from the published figures. *Error bars indicate the reading uncertainty of the digitization.*

We digitized the zero-field and saturated high-field resistance from the published parallel-field transport curves of Mertes et al.(*42*) (their Fig. 2) and Li et al.(*44*) (their Fig. 1), and formed $\rho_{sat}/\rho(0)$ and $R_{sat}/R(0)$, respectively, as functions of carrier density. Figure S11 shows that both ratios exhibit a pronounced maximum near the reported critical density and decrease toward both higher and lower densities. The maximum values are approximately 36 for Mertes et al. and 44 for Li et al. Near $n_c$ in bilayer $MoSe_2$, the resistance has already reached its high-field plateau by 11.8 T, so $R(11.8\ T)/R(0) \approx R_{sat}/R(0)$.

**Section 10: Estimate of gate-screening length scale**

Nearby metallic gates screen the long-range part of the Coulomb interaction and can reduce the stability range of a Wigner solid. A recent quantum Monte Carlo study of a disorder-free symmetric dual-gated two-dimensional electron gas found that Wigner crystallization is suppressed below a critical gate distance(*56*)

$$d_c \approx 29a_B^*,$$

where $d$ is the distance from the carrier layer to each metallic gate, measured in units of the effective Bohr radius $a_B^*$. Physically, this criterion corresponds to the spacer thickness below which metallic-gate screening makes the long-range Coulomb interaction too short-ranged to stabilize a Wigner crystal in the clean-limit calculation.

For an anisotropic dielectric environment such as hBN, the physical spacer thickness $t$ is rescaled as

$$d_{\text{eff}} = t\sqrt{\frac{\epsilon_{\text{in}}}{\epsilon_{\text{out}}}},$$

where $\epsilon_{\text{in}}$ and $\epsilon_{\text{out}}$ are the in-plane and out-of-plane hBN dielectric constants, respectively. We use representative values $\epsilon_{\text{in}} = 6.9$(*57*) and $\epsilon_{\text{out}} = 3.0$(*53*). Using the effective Bohr radii adopted in the main-text $r_s$ comparison,

$$a_{B,e}^* = 0.431 \text{ nm}, a_{B,h}^* = 0.185 \text{ nm},$$

the corresponding isotropic critical gate distances are

$$29a_{B,e}^* \approx 12.5 \text{ nm}, 29a_{B,h}^* \approx 5.4 \text{ nm}.$$

Applying the hBN anisotropy correction gives the physical spacer-thickness thresholds

$$t_{c,e} = \frac{29a_{B,e}^*}{\sqrt{\epsilon_{\text{in}}/\epsilon_{\text{out}}}} \approx 8.2 \text{ nm},$$
$$t_{c,h} = \frac{29a_{B,h}^*}{\sqrt{\epsilon_{\text{in}}/\epsilon_{\text{out}}}} \approx 3.5 \text{ nm}.$$

The present device has a 10-nm bottom hBN dielectric and a 13-nm top hBN dielectric. The published criterion assumes symmetric gates, whereas our device has unequal top and bottom spacer thicknesses. We therefore use the nearest gate distance, 10 nm, as the conservative comparison. This places the electron side close to, but above, the clean-limit gate-screening threshold, while the hole side is well above it. Using a symmetric effective distance based on the 10/13-nm stack would increase the effective gate distance and would not change this conclusion.

The criterion was obtained from a disorder-free symmetric dual-gated 2DEG calculation. In real devices, disorder can pin and locally stabilize Wigner-solid domains; the resulting thickness scale should therefore be viewed as a clean-limit gate-screening benchmark rather than as a hard boundary for the stability of the disorder-pinned Wigner solid in the device.